\documentclass[12pt,noinfoline]{imsart}

\usepackage{amsmath,amssymb,amsthm,amscd,amsfonts,bm,amsbsy}
\usepackage{graphicx}%
\usepackage{fancyhdr}
\usepackage{geometry}
\usepackage{url}
\usepackage{hyperref}
\usepackage{color}
\usepackage{natbib}
\usepackage{wrapfig,verbatim}
\usepackage{blkarray}
\usepackage{enumitem}
\usepackage{tikz-cd}
\usepackage{tikz}
\usepackage{mathtools}
\usepackage{dsfont}

\usepackage[ruled,vlined]{algorithm2e}

\theoremstyle{plain} 
\newtheorem{thm}{Theorem}[section]

\newtheorem{cor}[thm]{Corollary}

\newtheorem{proposition}[thm]{Proposition}
\newtheorem{prop}[thm]{Proposition}
\theoremstyle{definition}
\newtheorem{defn}[thm]{Definition}

\newtheorem{obs}[thm]{Observation}
\newtheorem{rmk}[thm]{Remark}
\newtheorem{example}[thm]{Example}

\newtheorem{problem}[thm]{Problem}

\newcommand{\R}{\mathbb{R}}

\newcommand{\N}{\mathbb{N}}
\newcommand{\Z}{\mathbb{Z}}

\newcommand{\minus}{\backslash}
\newcommand{\margin}{t}
\newcommand{\fiberofb}{\mathcal F(b)}

\begin{document}

\title{Markov bases: a 25 year update}

\author{\fnms{F\'elix Almendra-Hern\'andez}\thanksref{addr1}\corref{}\ead[label=e1]{falmendrahernandez@ucdavis.edu}},
\author{\fnms{Jes\'us A. De Loera}\thanksref{addr1}\ead[label=e2]{deloera@math.ucdavis.edu}},
\author{\fnms{Sonja Petrovi\'c}\thanksref{addr2}\ead[label=e3]{sonja.petrovic@iit.edu}}

\thankstext{addr1}{Department of Mathematics, University of California, Davis}
\thankstext{addr2}{Department of Applied Mathematics, Illinois Institute of Technology}

\date{\today}

\begin{abstract} In this paper, we evaluate the challenges and best practices associated with the Markov bases approach to sampling from conditional distributions. We provide insights and clarifications after 25 years of the publication of the Fundamental theorem for Markov bases by Diaconis and Sturmfels.
In addition to a literature review, we prove three new results on the complexity of Markov bases in hierarchical models, relaxations of the fibers in log-linear models, and limitations of partial sets of moves in providing an irreducible Markov chain.

\end{abstract} 

\maketitle

\section{Introduction} 

Algorithms for sampling from conditional distributions of discrete exponential families have a long history and a broad variety of applications. 
We investigate sampling from the conditional distributions given sufficient statistics in log-affine models, as defined in \cite{Lauritzen}, which are exponential family \cite{Barndorff-Nielsen} models  for $k$ discrete random variables $X_1,\dots,X_k$ with finite state spaces. 
A major application for constructing such a sample is to perform a goodness-of-fit test for the model on a large, but possibly sparse, state space. Typical uses are in the setting of networks or relational data described in \cite{FienbergWasserman1981categorical} in the context of contingency tables and sampling problems. Another use is for first- and higher-order Markov chains \citep{BesagMondal}. 

\citet{DS98} introduced a new sampling algorithm for this broad class of problems, showing  how to explicitly construct a Markov chain algorithm over the support of the conditional distribution given the log-linear exponential family's sufficient statistic.  
The crucial fact is that the chain is built using moves that have analogues in polynomial algebra and polyhedral geometry. Each instance inherits a finite set of moves necessary to obtain an irreducible chain for \emph{any} 
value of the sufficient statistic. They named any such finite collection of moves a \emph{Markov basis}.
In fact, the celebrated theorem that a set of \emph{moves} is a Markov basis if and only if the corresponding set of binomials generates the polynomial ideal of the model \cite[Theorem 3.1]{DS98} is often referred to as 
``\emph{The Fundamental theorem of Markov bases}"  in the   algebraic statistics community. 
Since their seminal paper, a flurry of research commenced, studying various aspects of Markov bases and 
their structure for special sets of discrete exponential families. While it is impossible to cite the entire bibliography, 
we recommend the comprehensive references \cite{AHT2012} and \cite{SethBook}. These books provide 
a compilation of the extensive work that has been conducted in this field so far.

In this paper we evaluate the challenges and best practices of this technique after 25 years of the birth of the Fundamental theorem of Markov bases. Before this theorem existed, there were a lot of concerns in the statistics community about how to produce an appropiate set of moves. For example, \cite{BesagClifford89} stated already over three decades ago, ``\textit{the MCMC algorithms have been successful in the sense that they have rejected hypothesized models with annoying regularity! Also the MCMC computations have confirmed the frequent irrelevance of standard asymptotic approximations in practice. Despite these comments, a practicable method of identifying irreducible sets of moves in constrained state spaces would be a considerable advance}".

Unfortunately, we do not believe that the Fundamental theorem of Markov bases is  well-known or properly understood. For instance, there is commentary in the literature that 
  the construction of algebraic Markov bases is `probably too theoretical' or `impractical in many applied settings'. One of the goals of this paper is to clarify the true limits of the Markov bases approach. We make a special effort to connect algebraic and polyhedral advances to classical statistics, particularly given explicit context. We provide new positive and negative results regarding Markov bases, incomplete sets of moves, and constrained state spaces.

Our first contribution is a review of the literature trying to clarify a few challenges and best uses of Markov bases. All clarifications and corrections of the existing literature will be stated as Propositions. Along the way, we will prove the following new results: 
\begin{enumerate}
\item There is no upper bound on the ``negative" relaxation of the fibers needed to connect the original fiber in general log-linear models; see Theorem \ref{thm:Unbounded_q}. This expands on previous work by Sturmfels (see \cite[Chapter 12]{St}). 
\item The existence of Markov bases elements arbitrarily complicated  holds even in the case of relaxed fibers: in Theorem \ref{thm:bad-news_restricted} and Proposition \ref{prop:anti-stair} we prove that relaxing a constraint set of entries $S$ still leads to complicated elements inside Markov bases when $S$ is chosen poorly. This significantly extends the negative results of \cite{DeLoera-Onn_all_linear}, \cite{slimTables}.
\item When working with hierarchical models on $d_1\times \cdots \times d_k$-tables, we bound the size of the 
Graver basis of the model by a polynomial on a proper subset of the levels $\{d_l\}_{l=1}^k$; see Corollary \ref{cor:polynomial_hierarchical}. 
\end{enumerate}

This paper is organized as follows: Markov bases are introduced in Definition~\ref{defn:Markov basis}, along with Proposition~\ref{prop:FTMB} that says they are  finite and that they provide connected chains for every log-linear model and every observed data table.   
Since Markov bases are insufficient for restricted fibers, and one may need arbitrarily large set of moves, Proposition~\ref{thm:bounded graver is markov for bounded fibers} provides bases for all constrained state 
spaces for all log-linear models and every observed data table. 
In Section \ref{sec:how_complicated} we summarize known negative and positive results about Markov basis complexity.  In particular, Markov bases can be arbitrarily complicated already for the simplest non-decomposable hierarchical model. However, when the model possesses a specific block structure and sparsity patterns, we are able to recover Proposition \ref{prop:graver_n_fold} and prove Corollary \ref{cor:polynomial_hierarchical}, a polynomial bound on the size of bases needed to connect restricted fibers.  

Proposition~\ref{prop:existence_q} explains precisely why special basis elements, such as lattice bases, or sets of basic moves, do not provide an irreducible Markov chain in general. 
In Section~\ref{sec:relaxation} we prove that the approach, proposed by many authors, of relaxing table entries to allow negative entries cannot work in general. The new results will be stated in Corollary~\ref{cor:polynomial_hierarchical}, Theorems~\ref{thm:Unbounded_q} and \ref{thm:bad-news_restricted}, and Proposition~\ref{prop:anti-stair}.
All proofs appear in the Appendix \ref{appendix:proofs}.

Code for computing the examples in this paper is provided on the following GitHub repository:
 \href{https://github.com/Sondzus/MarkovBases25years}{https://github.com/Sondzus/MarkovBases25years}.

\section{Log-linear models, fibers, and Markov bases}\label{sec:definitions}

Denote by $X=(X_1,\dots,X_k)$ a discrete random vector in the state space $\mathcal X=[d_1]\times\dots\times[d_k]$, where $[d]=\{1, 2, \ldots, d\}$ for any positive integer $d$. The \emph{log-affine} model is the discrete exponential family 
$$f_\theta(x)= h(x) e^{\eta(\theta)^T \margin(x)-\psi(\theta)},$$
where the sufficient {statistic} is $\margin(x):\mathcal X\to\mathbb R^m$ , {natural parameter} $\eta(\theta):\Theta\to\mathbb R^m$, and log-partition function  
$\psi:\Theta\to\mathbb R$. Here $T$ denotes vector transpose, so that the product above is the usual inner product. 
From the point of view of algebra and geometry, one may assume that $h(x)=\bf 1$ is a constant 1 vector; in this case the model is \emph{log-linear}. This assumption simplifies the algebraic considerations below, and can easily be rectified by a substitution of indeterminates to recover the original log-affine model. 

The model representation of interest to us arises when data are arranged in a contingency table cross-classifying items according to the $k$ categories,  so that   the sufficient statistic computation  is realized as a linear operation on the table. Namely, consider $U\in \mathbb Z^{d_1\times\dots\times d_k}_{\geq 0}$, a $k$-dimensional table  
of format  $d_1\times\dots\times d_k$, whose sample space $\mathbb Z^{d_1\times\dots\times d_k}_{\geq 0}$ naturally corresponds to $\mathcal X$.  In the table $U$,  the $(i_1,\dots,i_k)$-entry $u_{i_1,\dots,i_k}$ records the number of instances of the joint event $(X_i=i_1,\dots,X_k=i_k)$ in the data.  There is a corresponding probability table  $P$, of the same format $d_1\times\dots\times d_k$, whose cells  $p_{i_1,\dots,i_k}$ are joint probabilities. 
Algebraically, since computing marginals $\margin(U)$ is a linear  operation on cells of the table $U$, one can also simply write $\margin(U)=AU$ for some integer matrix $A$ and $U$ written in vectorized form $X\in\mathbb Z^{d_1\cdots d_k}_{\geq 0}$. We use the same notation $U$ for both the matrix and table format, as the entries are identical, the vector format is just a matter of flattening the table into a vector. 
A log-linear model for $X_1,\dots,X_k$ can be equivalently expressed as 
\begin{equation}\label{log-linear model}
	P(X=x) = P(U=u) \propto \exp\{ \left<Au,\theta\right>\}, 
\end{equation}
whose sufficient statistics $t(x)$ are some set of \emph{marginals} $Au$ of the contingency table. 
Geometrically, the model is completely determined by the \emph{design matrix} $A$. 

\begin{example}\label{example: job data}
	Consider the following data on job satisfaction from \cite[p.57]{Agresti2002}; this example is also used in the documentation page for the R package {\tt algstat} \citep{algstat.R}. 
\begin{table}[ht]
\centering
\begin{tabular}{r|rrrr|r}
 & VeryD & LittleD & ModerateS & VeryS \\ 
  \hline
$<$ 15k & 1 & 3 & 10 & 6  & 20 \\ 
  15-25k & 2 & 3 & 10 & 7  & 22 \\ 
  25-40k & 1 & 6 & 14 & 12  & 33 \\ 
  $>$ 40k & 0 & 1 & 9 & 11  &21\\ 
   \hline
   &4&13&43&36& \\
\end{tabular}
\end{table}

The table classifies individuals according to $k=2$ categorical variables: $d_1=4$ levels of salary ranges and $d_2=4$ levels of job satisfaction/dissatisfaction. The table $U$ is thus of size $4\times4$, and can be flattened into a vector of length $16$: $u=[1,2,1,0, 3,3,6,1,10,10,14,9,6,7,12,11]$.

For the model of independence $X_1\perp \!\!\! \perp X_2$, marginals are row and column sums, and thus the equation $Au = t(u)$ is written using the following design matrix: 
\tiny
\[ A = \left[ 
\begin{tabular}{rrrrrrrrrrrrrrrrr}
1 & 1 & 1 & 1 & 0 & 0 & 0 & 0 & 0 & 0 & 0 & 0 & 0 & 0 & 0 & 0 \\ 
0 & 0 & 0 & 0 & 1 & 1 & 1 & 1 & 0 & 0 & 0 & 0 & 0 & 0 & 0 & 0 \\ 
 0 & 0 & 0 & 0 & 0 & 0 & 0 & 0 & 1 & 1 & 1 & 1 & 0 & 0 & 0 & 0 \\ 
 0 & 0 & 0 & 0 & 0 & 0 & 0 & 0 & 0 & 0 & 0 & 0 & 1 & 1 & 1 & 1 \\ 
1 & 0 & 0 & 0 & 1 & 0 & 0 & 0 & 1 & 0 & 0 & 0 & 1 & 0 & 0 & 0 \\ 
 0 & 1 & 0 & 0 & 0 & 1 & 0 & 0 & 0 & 1 & 0 & 0 & 0 & 1 & 0 & 0 \\ 
0 & 0 & 1 & 0 & 0 & 0 & 1 & 0 & 0 & 0 & 1 & 0 & 0 & 0 & 1 & 0 \\ 
0 & 0 & 0 & 1 & 0 & 0 & 0 & 1 & 0 & 0 & 0 & 1 & 0 & 0 & 0 & 1 \\ 
\end{tabular}
\right]\text{\normalsize .}
\] 
\normalsize
One easily checks that 
\[A\cdot 
[1,2,1,0, 3,3,6,1,10,10,14,9,6,7,12,11]^T  =  [4,   13,   43,  36,  20,   22,  33,  21],\]
 providing the above table marginals. 
 In anticipation of Definition~\ref{defn:Markov basis}, note that the table $b=[1,-1,0,0, -1,1,0,0, 0,0,0,0, 0,0,0,0]$ satisfies $Ab=0$. In particular, adding $b$ to the observed table $u$ does not change the table margins.  
 \end{example}

\begin{defn}\label{defn:fiber}
A \emph{fiber} of the observed table $u$ under the log-linear model defined by $A$ is the set of all integer tables  with the same values of the sufficient statistics. One can write it in both table and vector format as follows: 
$$\fiberofb = \{u\in\mathbb Z^{d_1\times\dots\times d_k}_{\geq 0}: \margin(u)={b}\}
\equiv \{u\in\mathbb Z^{d_1\cdots d_k}_{\geq 0}: Au={ b}\}.
$$ 
\end{defn}

Let us reflect on some implications of Definition~\ref{defn:fiber}: 
\begin{enumerate}
	\item The fiber $\fiberofb$ depends on the model, so really the fiber notation should include the matrix $A$. However, for readability, we suppress the dependence on the matrix $A$ from the notation whenever it is clear from the context. 
	\item The fiber $\fiberofb$ is the support of the conditional distribution given the observed value of the sufficient statistic $\margin(u)$. Therefore, it is critical to exact conditional testing. 
	\item The fiber $\fiberofb$, as defined, is unconstrained above, except by the total sample size.
\end{enumerate} 
The first  point is technical. The second point is the most compelling reason for defining the fiber. The last point, however, raises questions that have not yet been answered {in general} in the statistics literature. These are detailed in Section~\ref{sec:constraints}.

\begin{example}[\ref{example: job data}, continued]\label{example:job data fiber}
In the job satisfaction example, the fiber is the set of all $4\times 4$ nonnegative integer tables whose column sums are $[4,   13,   43,  36]$ and whose row sums are $[20,   22,  33,  21]$. The fiber size for this data is $90,208,550$ tables. In other words, $|\mathcal F\big([4,   13,   43,  36,\allowbreak 20,   22,  33,  21]\big)|=90,208,550$.
\end{example}

\begin{example}\label{ex:no3way} The \emph{no-three-way interaction model} is the family of probabilities $(p_{i_1, i_2, i_3})$ on $I\times J\times K$ tables such that 
\[
\log p_{i_1, i_2, i_3}= \mu+\mu_{12}(i_1, i_2)+\mu_{23}(i_2, i_3)+\mu_{13}(i_1, i_3),
\]
where $\mu_{12}, \mu_{13}$ and $\mu_{23}$ are free vectors of parameters and $\mu$ is a normalizing constant. 
This is a log-linear model whose design matrix $A$ is such that for every $I\times J\times K$ contigency table $u$,  
$Au=(u_{12}, u_{23}, u_{13})$ where 
\[
u_{12}(i_1, i_2)=\sum_{k=1}^Ku_{i_1, i_2,k},\;\;\;\;\; u_{23}(i_2, i_3)=\sum_{i=1}^I u_{i, i_2, i_3}, \;\;\;\text{and }\;\; u_{13}(i_1, i_3)=\sum_{j=1}^J u_{i_1, j, i_3}.
\]
The significance of this model is that \emph{any} possible fiber of \emph{any} log-linear model is linearly isomorphic to a fiber of the no-three-way interaction model on $I\times J\times 3$ tables for appropriate $I$ and $J$. This will be explained in Proposition \ref{prop:all-linear-are-transportation} below. In other words, this model is \emph{universal}, in the sense of polyhedral geometry.  
\end{example}

To sample from model fibers using a Markov chain, one needs to define a basis for the chain. For example, such a Markov basis for the no-3-way interaction model from~\ref{ex:no3way} is provided in \cite[p.379]{DS98} in terms of constant log-odds interpreted as binomial relations on the cells of the  $3$-way  joint probability table.  
This construction generalizes. 

The key to defining a Markov basis is to realize that for any two tables $u,v\in\fiberofb$ in the same fiber, their entry-wise difference is a zero-margin table, so that  $\margin(u-v)=0$. In matrix-vector format, $A(u-v)=0$. That is, the vector $u-v$ is in the integer kernel of the matrix $A$, denoted by $\ker_\mathbb Z A\subset \Z^{d_1\cdots d_k}$.  Conversely, each vector in the kernel can be written as a difference of two non-negative vectors. 
\begin{defn}[Markov basis]\label{defn:Markov basis}
	Fix the exponential family model of $k$-way tables with design matrix $A$ and sufficient statistics $Au$. 
	 Any zero-margin table $b\in \ker_\mathbb Z A$ is called a \emph{move}. 
	A set $\mathcal M\subset \ker_\mathbb Z A$ of zero-margin tables  is a \emph{Markov basis} for the model 
	 if for every value of the marginal vector $b$, $\mathcal M$ connects any pair $u, v\in \fiberofb$ through points inside $\fiberofb$. 
	 In other words, there exists a choice of $L$ elements $m_1,\dots,m_L\in\pm \mathcal M$  such that $u= v + \sum_{i=1}^L m_i$, 
	with the requirement that  $u=v + \sum_{i=1}^\ell m_i\in\fiberofb$ for all $\ell\leq L$.
\end{defn} 
That an \emph{infinite} set of moves that connects all tables exists is trivial: every difference $u-v$ of two tables with same marginals is a feasible move. The question is how to remove redundant moves from this set to obtain a \emph{finite} basis. 
The following key result, the `\emph{Fundamental theorem of Markov bases}', holds for unconstrained fibers as defined in \ref{defn:fiber}. 

\begin{thm}[\textbf{Fundamental theorem of Markov bases}, \cite{DS98}, Thm. 3.1]\label{prop:FTMB}
The set $\mathcal M\subset \ker_\mathbb Z A$  is a Markov basis for $A$ if and only if the corresponding set of binomials generates the toric ideal $I_A$.
\end{thm}

\begin{example}[\ref{example: job data}, continued]\label{example:job data binomials}
Recall the  vector $b=[1,-1,0,0, -1,1,0,0, 0,0,0,0, 0,0,0,0]\in\ker_\mathbb ZA\subset \mathbb Z^{16}$  for the design matrix $A$ in the previous example of the model of independence. In table form, this vector represents: 
\begin{table}[ht]
\centering
$b = $
\begin{tabular}{r|rrrr|r}
 & VeryD & LittleD & ModerateS & VeryS \\ 
  \hline
$<$ 15k & 1 & -1 & 0& 0 & 0 \\ 
  15-25k & -1 & 1 & 0 & 0  & 0 \\ 
  25-40k & 0 & 0 & 0& 0  & 0 \\ 
  $>$ 40k & 0 & 0 & 0& 0  &0\\ 
   \hline
   &0&0&0&0& \\
\end{tabular}.
\end{table}

\noindent 
The zero-margin table $b$ is thus equivalent to  the familiar cross-product ratio $\frac{p_{11}p_{22}}{p_{12}p_{21}}=1$, which can algebraically be represented by a polynomial $p_{11}p_{22}-p_{12}p_{21}=0$. 
We think of this binomial as a `relation' on the joint cell probabilities. 
The toric ideal $I_A$ contains many more such binomials; a Markov basis is a generating set sufficient to derive \emph{all} binomial relations which hold for the given model. 
\end{example}

We will not delve further on the algebraic details, except to note that the algebra correspondence \emph{guarantees} that a {finite} Markov basis exists for every log-linear model, because by the Hilbert basis theorem each ideal of a polynomial ring has a finite generating set \cite[p.76]{CLO}.  Additional information and the algebraic perspective background can be found in \cite{DSS09};  see also \cite{AHT2012} for a comprehensive review.

\begin{rmk}
It is important to be careful when considering discrete exponential families as log-linear models. Referring back to Equation~\eqref{log-linear model}, and the definitions preceding it on page \pageref{log-linear model}, for the Fundamental theorem of Markov bases to be useful, the sufficient statistic $t$ must be a linear function of the data. 
However, when $t$ is not linear on $u$, the canonical way  of representing a discrete exponential family as a log-linear model is not always useful.  Specifically, the canonical log-linear form  from  \cite{GeigerMeekSturmfels} first identifies the sample space with unit vectors by representing any table $u$ with its corresponding standard basis vector $\bm e_u\in \R^{|\mathcal T(N)|}$, where $\mathcal T(N)=\{u\in\mathbb Z^{d_1\cdots d_k}_{\geq 0}: ||u||_1=N\}$. Second, it defines  the  $1\times |\mathcal T(N)|$ design matrix $\tilde A$, letting the $u$-th entry of $\tilde A$ be equal to $t(u)$. 
The exponential family parametrization then becomes log-linear in form $f_\theta(u)=\propto\exp\{\langle t(u), \theta\rangle\} =\exp\{\langle \tilde A\bm e_u, \theta\rangle\}$. This suggests that we can apply Markov basis theory to the model $\mathcal M_\Theta$ using the $\tilde A$ as the design matrix. 
The fibers of the model, by definition, $\mathcal F(b)=\{u\in\Z^{d_1\times \cdots\times d_k}_{\geq 0}: t(u)=b\}\equiv \{\bm e_u: \tilde A\bm e_u=b\}\not\equiv \{y\in \Z_{\geq 0}^{|\mathcal T(N)|}:\tilde Ay=b\}$ are no longer described as the integer points of a polytopal region. Given that the set $\mathcal F(b)\equiv \{\bm e_u: \tilde A\bm e_u=b\}$ is linearly independent, it follows that at least $|\mathcal F(b)|-1$ vectors are needed to connect all the points inside the fiber. This is equivalent to enumerating the points of $\mathcal F(b)$, which is infeasible in practice and $\#$P-hard in terms of theoretical complexity. Further study of this issue will appear in forthcoming work, while specifics of the representational difficulty for discrete exponential families for random graphs appear in \cite{GPS21+}. 

The \emph{Ising model} presented in \cite[Section 2]{DelCampoCepedaUhler2017} serves as a compelling example of a model that cannot be analyzed using Markov basis theory.  In their work, the authors 
 treat the Ising model as a log-linear model by implicitly introducing the matrix $\tilde A$ and claiming that finding a Markov basis for the Ising model is ``computationally intractable." However, the Ising model does not have a linear sufficient statistic with respect to its parameters, rendering the application of Markov basis theory to their model description useless.
 Nonetheless, the authors are able to exploit the combinatorial structure of the Ising model to devise an exact goodness-of-fit test without using Markov basis theory. This underscores the importance of understanding the assumptions underlying the use of Markov basis theory and highlights the need for alternative methods when these assumptions are not met.

\end{rmk}

\section{Myths and issues with Markov Bases} 

As we will see in Proposition \ref{cor:bad-news}, Markov bases, although well-defined and finite, can exhibit significant complexity. Their structure can vary, and computation times can become practically infinite. In statistics, this translates to moves so large in norm that one cannot hope to compute a minimal set of moves necessary for sampling all the model fibers. The issue is compounded by the fact that  most of the moves generated by the algebraic construction in Proposition~\ref{prop:FTMB} are not useful for a given data set because Markov bases are data agnostic; 
 cf. \cite{SteveAleMe-holland}, \cite[Problem 5.5]{DobraEtAl-IMA}. 
While some of the early works have acknowledged this issue,  practical implementations that only construct `useful' moves are not many. One line of work to address this problem began with \cite{PRF:09} supporting  network modeling; another  with \cite{Dob2012} who computes local Markov moves dynamically. 

Nevertheless, the theory remains: it is  {completely clear} by  Theorem~\ref{prop:FTMB} that the moves in a Markov basis  do connect the fiber: every fiber, for any observed data table, and every log-linear model.    
  This important fact should not be overlooked, and it was also spelled out quite clearly in \cite{BesagClifford89}: Markov chains for sampling from the conditional distribution on a fiber built using Markov bases are, by definition, irreducible. 

 Unfortunately, sometimes  statements about what Markov bases can and cannot do are misleading when taken out of context.  
  For example, \cite{ZhangChenJASA13} quote \cite{goldenberg2010survey} that \emph{``it is unclear whether the proposals in this literature are in fact reaching all possible tables associated with the distribution."}
While the authors are pointing to apparent open problems in the context of network models, this statement does not hold for general unconstrained models on contingency tables. 
Indeed, log-linear models arising in the context of random networks do  pose additional challenges including sparsity and sampling constraints. Even so, Proposition~\ref{thm:bounded graver is markov for bounded fibers} solves the issue raised by the quote above. 

In the remainder of this section, we reflect on the possibilities and difficulties that arise in practice, and related literature on challenges related to fiber sampling.

\subsection{The effect of sampling constraints and structural zeros}  
\label{sec:constraints}

Fibers $\fiberofb$, as defined, are unconstrained, except by the total sample size, meaning that they consist of \emph{any} possible observable contingency table with the given marginal counts. In practice, it is often the case that some or all of the cells of the table are bounded. There are at least three common scenarios. First, cells may be bounded by sampling constraints that are lower than the marginal bound; for example in ecological inference through individual-level information. \cite[Section 2]{RPF:11} 
describes the generalized $\beta$ model for random graphs, in which 
data are represented by $n\times n$ contingency table and each cell $U_{ij}$ takes values in $\{0,\dots,N_{ij}\}$ for deterministic positive integers $N_{ij}$; compare to the \emph{Rasch model} (\cite{Rasch}, \cite{Andersen}) which is equivalent to a bipartite graph and ``\textit{is concerned with modeling the joint probabilities that $k$ subjects provide correct answers to a set of $l$ items, and is one of the most popular statistical models used in item response theory and in educational tests}".  
The second scenario is when cell bounds arise from disclosure limitation problems; see \cite[p.2]{Dob03} and discussion therein. Third, the model may contain structural zeros as discussed in \cite[Ch. 8]{CategoricalDataAnalysis}.

A critical failure of Markov basis from Definition~\ref{defn:Markov basis} is they do not connect  fibers restricted by cell bounds.  In practice, this issue is generally solved on a case-by-case basis. 
 \cite{RY10}  spell out various scenarios for having cell bounds or structural zeros and offer partial solutions. Sequential importance sampling algorithms for $2$-way tables with $0/1$ entries and structural zeros are provided in \cite{Chen-2way-zeros}. 
To quote \cite{SteveAleMe-holland},  ``\emph{Markov bases we obtain from [the Theorem~\ref{prop:FTMB}] violate the $p_1$ [random graph model] constraints that each dyad [of the random graph] is associated with a multinomial of size $1$}'' (i.e., the constraint that for each dyad we observe one and only one of the possible dyadic configurations). 
In the context of random graph log-linear exponential family models, which are $0/1$ contingency tables, explicit bases derivations exist for the $\beta$-model of simple graphs \cite{OHT:11} and  $p_1$ model of directed graphs \cite{GPS21+}.  They are based on  \cite[Proposition 2.1]{HT:10} which states that a 0/1 subset of the Graver basis connect 0/1 fibers. 
A general version of this result is the answer to sampling restricted fibers. 

\begin{defn}
Let $q \in\mathbb Z_{\geq 0}$ be the sampling constraint bound on table entries. Define the $q$-bounded fiber as $\mathcal F^q(b):=\{u\in\mathbb Z^{d_1\times\dots\times d_k}: \margin(u)={b} \mbox{ and } q\geq u_{i_1\dots i_k}\geq 0\}$. 
\end{defn} 

We now introduce a superset of moves that will allow us to connect $q$-bounded fibers.

\begin{defn}[Graver basis]
The \emph{Graver basis} $Gr(A)$ of $A$ is the set of minimal elements in $\ker_\Z A\minus \{0\}$ with respect to the well partial order defined by $x\sqsubseteq y$ when $|x_i|\leq |y_i|$ and $x_iy_i\geq 0$ for all $i$.
\end{defn}  

Note that each Graver element is just another fiber move on the table: these vectors correspond to tables whose $A$-margins are zero. Furthermore, it follows that $Gr(A)$ contains any minimal Markov basis for the log-linear model with design matrix $A$. It is important to note that the Graver basis is typically much larger than a minimal Markov basis. In other words, it contains more moves than necessary to sample from unconstrained fibers. 
 The following result states that, in fact, $Gr(A)$ contains enough moves to connect \emph{any restricted fiber}.   

\begin{prop}\label{thm:bounded graver is markov for bounded fibers}
Let $b$ be any value of the $A$-margins and let $q\in\mathbb Z_{\geq 0}$ be the sampling constraint bound on table entries. The restricted fiber   $\mathcal F^q(b)$ is connected by the subset $G\subset Gr(A)$ whose entries are each bounded by $q$. 
\end{prop} 

The proof  can be found in Appendix \ref{appendix:proofs}. 
This  has also appeared recently as \cite[Theorem 9.4.5]{SethBook}, and it has been known for some time in algebraic literature as a straightforward generalization of the 0/1 table result \cite[Theorem 2.1]{HT:10}.

\subsection{How complicated are Markov bases in general?}\label{sec:how_complicated}

In this section we provide a compilation of results regarding the complexity of Markov basis. To do so, we will focus on a special type of log-linear models whose marginals are described by groups of interaction factors among the random variables $X_1, \ldots, X_k$ that follow the familiar hierarchy principle (see, for example, \cite{CategoricalDataAnalysis}), which postulates that if an interaction term is zero, then so are any of the higher-order terms containing it. Combinatorially, such a collection of marginals can be succinctly summarized by a \emph{simplicial complex}, or an abstract simplicial complex, defined as a family of subsets of the ground set $\{1,\dots,k\}$ that is closed under taking subsets. 
A \emph{hierarchical model} is thus described by a list of maximal---maximal under containment---faces $F_1, \ldots, F_r$ of a simplicial complex $\Delta$ with ground set $[k]$ and levels $d_1, d_2, \ldots, d_k$. The design matrix $A_\Delta$ of this model induces a linear map from $\Z_{\geq 0}^{D}$ to $\Z_{\geq 0}^{D_1}\oplus \cdots\oplus \Z_{\geq 0}^{D_r}$, with $D=d_1\cdots d_k$ and $D_s=\prod_{i\in F_s}d_i$. The map is such that for every $u=u_{i_1, \ldots, i_k} \in \Z_{\geq 0}^{D}$, 
\[
A_\Delta u = (u_{F_1}, u_{F_2}, \ldots, u_{F_r})\;\;\;\;\text{where }\;\; u_{F_s}=\sum_{(i_j:j\in [k]\minus F_s)} u_{i_1, \ldots, i_k}.
\]
Notice that for a fixed $s$, the sum in $u_{F_s}$ is being taken over all the elements of $\prod_{j\in [k]\minus F_s}[d_j]$.  
A simplicial complex $\Delta$ is \emph{reducible} with decomposition $(\Delta_1, S, \Delta_2)$ and \emph{separator} $S\subset |\Delta|$ if $\Delta=\Delta_1 \cup \Delta_2$ and $\Delta_1\cap \Delta_2=2^S$, where $|\Delta|=\bigcup_{F\in \Delta}F$ and $2^S$ is the power set of $S$. A simplicial complex $\Delta$ is \emph{decomposable} if it is reducible and $\Delta_1$, $\Delta_2$ are decomposable or if they are of the form $2^R$ for some $R\subset |\Delta|$.

On one hand, it is known that for contingency table hierarchical models that are decomposable, the structure of their Markov basis is well understood thanks to \cite{takken2000} and \cite{Dob03}. In fact, one can always find a Markov basis with moves whose one-norms equal 4, regardless of the size of the levels $d_1, \ldots, d_k$. This was subsequently used to spell out a divide-and-conquer algorithm to compute Markov bases for reducible models in \cite{DS04}. In practice, this translates to \emph{scalability} of Markov bases for exact conditional testing.

On the other hand, such a bound fails to exist for even the simplest non-decomposable model: the  no-three-way interaction of three discrete random variables  \citep{CategoricalDataAnalysis} described in Example \ref{ex:no3way}. By importing powerful polyhedral geometry results into statistics, \cite{slimTables} showed that the minimal Markov bases of the no-three-way-interaction model on $I\times J\times 3$ tables can contain moves with arbitrarily large 1-norm, if $I$ and $J$ are unrestricted.

Before explicitly stating the result, we introduce some notation. For a matrix $A$ and a marginal vector $b$, we define the \emph{polytope} $P_{A, b}=\{x\in \R^{d_1\times \cdots\times d_k}:Ax=b, x\geq 0\}$ as the space of solutions of the linear equations $Ax=b$, bounded by the halfspaces $H_{i_1,\ldots, i_k}:=\{x\in \R^{d_1\times\cdots\times d_k}: x_{i_1, \dots, i_k}\geq 0\}$ for every $(i_1, \ldots, i_k)\in [d_1]\times \cdots\times [d_k]$. From this, we observe that $\mathcal F(b)=P_{A, b}\cap \Z^{d_1\times \cdots\times d_k}$, which means that the fiber $\mathcal F(b)$ corresponds to the set of integer points inside $P_{A,b}$. The following remarkable result shows that the no-three-way interaction model can capture the geometric structure of any polytope $P_{A, b}$.

\begin{prop}[\cite{DeLoera-Onn_all_linear}]\label{prop:all-linear-are-transportation}
For any rational matrix $A$ and any integer marginal vector $b$, $P_{A,b}=\{y\in \R^{n}_{\geq 0}: Ay= b\}$ is polynomial-time representable as a slim $3$-way transportation polytope:
\[
T = \left\{x\in \R^{I\times J\times 3}_{\geq 0}: \sum_kx_{i,j,k}=u_{i,j}, \sum_jx_{i,j,k}=v_{i,k}, \sum_ix_{i,j,k}=w_{j,k}\right\}.
\]
For positive integers $h$ and $h'$, saying that a polytope $P \subset \R^{h}$ is representable as a polytope $Q \subset \R^{h'}$ means that there is an injection $\sigma: \{1, \ldots, h\}\to \{1, \ldots, h'\}$ such that the coordinate-erasing projection
\[
\pi: \R^{h'}\to \R^{h}: x=(x_1, \ldots, x_{h'})\mapsto \pi(x)=(x_{\sigma(1)}, \ldots, x_{\sigma(h)})
\]
provides a bijection between $Q$ and $P$ and between their integer points $Q\cap \Z^{h'}$ and $P\cap \Z^h$.
\end{prop}

As a consequence of Proposition \ref{prop:all-linear-are-transportation}, we have the following.  

\begin{cor}[\cite{slimTables}]\label{cor:bad-news} 
For any nonnegative integer vector $\theta\in \mathbb{N}^\eta$ there exists $I,J>0$ such that any Markov basis for the no-three-way interaction model on $I\times J\times 3$ tables must contain an element whose restriction to some $\eta$ entires is precisely $\theta$. In particular the degree and support of elements in the minimal Markov bases when $I$ and $J$ vary can be arbitrarily large.
\end{cor}

Despite the previous findings, we can still derive good complexity results for non-decomposable models as demonstrated in Corollary \ref{cor:polynomial_hierarchical} below. The result builds upon two key points:
\begin{enumerate}
\item often the design matrix of a hierarchical model exhibits a  block structure. For instance it could be an \emph{$n$-fold matrix}, defined below; and 
\item  the Graver basis of an $n$-fold matrix solely depends on its constituent blocks. 
\end{enumerate} 

\begin{defn}
	Given fixed matrices $A\in \Z^{p\times s}$ and $B\in \Z^{p' \times s}$ with positive integer $p,p',s$, the \emph{$n$-fold matrix} of the ordered pair $(A,B)$ is defined as the $(np+p')\times sn$ matrix 
\[
[A,B]^{(n)}:=
\begin{pmatrix}
	A & 0 & 0 & \cdots & 0\\ 
	0 & A & 0 & \cdots & 0\\ 
	\vdots & \vdots & \ddots & \vdots & \vdots\\
	0 & 0 & 0 & \cdots & A\\
	B & B & B & \cdots & B
\end{pmatrix}. 
\]   
\end{defn}

We define the type of a vector $x=(x^{(1)}, \ldots, x^{(n)})\in \Z^{sn}$ as the number $|\{j: x^{(j)}\neq 0\}|$ of nonzero components $x^{(j)}\in \N^s$. The following result establishes a stabilization property of the Graver basis for $n$-fold matrices.

\begin{prop}[\cite{HoSu}]
	Given matrices $A\in \Z^{p\times s}, B\in \Z^{p' \times s}$, there exists a constant $C$ such that for all $n$, the Graver basis of $[A,B]^{(n)}$ consists of vectors of type at most $C$. The smallest of these constants is known as the {Graver complexity} of $A,B$ and we denote it by $g(A,B)$. Furthermore, \[g(A, B)=\max_{x\in Gr(B\cdot Gr(A))}||x||_1.\]
\end{prop}

The Graver basis $Gr([A,B]^{(n)})$ for any $n$-fold of $A,B$ can be obtained from this result. 

\begin{prop}[\cite{N-fold2008}]\label{prop:graver_n_fold}
For fixed matrices $A\in \Z^{p\times s}$ and $B\in \Z^{p' \times s}$, the Graver basis $Gr([A,B]^{(n)})$ can be computed in polynomial time on $n$. Moreover, the size of $Gr([A,B]^{(n)})$ is bounded by $|Gr([A,B]^{(g)})|\binom{n}{g}$, where $g=g(A,B)$ is the Graver complexity of $A,B$.
\end{prop}

The following complexity result is  a bound on the size of the Graver basis of hierarchical log-linear models. In light of Proposition~\ref{thm:bounded graver is markov for bounded fibers}, which states that Graver elements contain all moves necessary for sampling contingency tables with constraints on the cell entries, it has direct implications on the feasibility of sampling restricted fibers.  
\begin{cor}\label{cor:polynomial_hierarchical} Let $\Delta$ be a simplicial complex with ground set $[k]$ and maximal faces $F_1, \ldots, F_r$. Let $V\subset [k]$ be such that for every $j\in [k]$, $V\subset F_j$ or $V\subset F_j^c$. Let $\delta=(\delta_l)_{l\not\in V}$ be fixed. Then, for any $(d_1, \ldots, d_n)\in \N^n$ with $(d_l)_{l\not\in V}=\delta$, the size of the Graver basis  $|Gr(A_\Delta)|$ is bounded by a polynomial in $\prod_{l\in V}d_l$.
\end{cor}

The proof of this corollary relies on the fact that $A_{\Delta}$ is an $\left(\prod_{l\in V}d_l\right)$-fold matrix, which allows us to directly apply Proposition \ref{prop:graver_n_fold}. In Appendix \ref{appendix:proofs}, we provide a proof of the corollary based on the idea presented in \cite[Lemma 2.2]{HoSu}.\\

\noindent
\begin{minipage}[h]{0.8\textwidth}
\begin{example} 
Let $\Delta$ be a simplicial complex on 4 vertices with levels $(d_1, d_2, 2, 3)$. The maximal faces of $\Delta$ are $F_1=\{1,2,3\}$, $F_2=\{1,2,4\}$, and $F_3=\{3,4\}$. The set $V=\{1,2\}$ satisfies $V\subset F_1, F_2$, and $V\subset F_3^c$. By the proof of Corollary \ref{cor:polynomial_hierarchical}, it follows that $A_{\Delta}=[A, B]^{(d_1d_2)}$, where $B=I_{6}$ and $A$ is the design matrix of the independence model with levels $(2,3)$. Using the software \verb|4ti2| from \cite{4ti2}, we can compute $g(A, B)=3$ and $|Gr([A,B]^{(3)})|=15$. Therefore, we have $|Gr(A_{\Delta})|\leq 15\binom{d_1d_2}{3}$ for any $d_1, d_2$.
\end{example}
\end{minipage}%
\begin{minipage}[h]{0.25\textwidth}
\centering
\begin{tikzpicture}
\node[circle, draw, fill=black, inner sep=1.5pt, label=above:1] (v1) at (-4/3,0) {};
\node[circle, draw, fill=black, inner sep=1.5pt,label=right:2] (v2) at (0,0) {};
\node[circle, draw, fill=black, inner sep=1.5pt, label= right :3] (v3) at (3/4,1) {};
\node[circle, draw, fill=black, inner sep=1.5pt, label= right:4] (v4) at (3/4,-1) {};

\fill[color=lightgray] (v1.center) -- (v2.center) -- (v3.center) -- cycle;
\fill[color=lightgray] (v1.center) -- (v2.center) -- (v4.center) -- cycle;

\draw (v1) -- (v2);
\draw (v1) -- (v3);
\draw (v1) -- (v4);
\draw (v2) -- (v3);
\draw (v2) -- (v4);
\draw (v3) -- (v4);

\node[circle, draw, fill=black, inner sep=1.5pt] at (-4/3,0) {};
\node[circle, draw, fill=black, inner sep=1.5pt] at (0,0) {};
\node[circle, draw, fill=black, inner sep=1.5pt] at (3/4,1) {};
\node[circle, draw, fill=black, inner sep=1.5pt] at (3/4,-1) {};

\end{tikzpicture}
\end{minipage}\\

\begin{rmk} An important assumption made in Proposition \ref{prop:graver_n_fold} is that the matrices $A,B$ are fixed. However, it is worth noting that the Graver complexity $g(A,B)$ can become arbitrarily large when the size of the entries in $A$ and $B$, or the dimensions of $A$ and $B$, vary. For example, it was recently shown in \cite[Theorem 5.1]{graver_complexity_onn2022} that $2\times 4$ matrices have arbitrarily large Graver complexity when we let the entries vary. Similar results can be also found in \cite{graver_complexity_onn2009}. 	
\end{rmk}

\begin{rmk} In scenarios where it is not feasible to explicitly compute the Graver complexity $g(A,B)$ for fixed matrices $A\in \Z^{p\times s}$ and $B\in \Z^{p'\times s}$, we can rely on upper bounds. The best known upper bounds come from recent developments on sparse integer programming where the tree-depth plays an important role (see \cite{graver_complexity_onn2022}, \cite{treedepth2022}, \cite{Knop_td_bound2020}). To clarify this point we introduce some definitions. The \emph{height} of a rooted tree refers to the maximum number of vertices on a path from the root to a leaf. For a given graph $G = (V, E)$, a rooted tree on $V$ is considered \emph{valid} for $G$ if, for each edge $\{j, k\} \in E$, either $j$ or $k$ lies on the path from the root to the other vertex. The \emph{tree-depth} $td(G)$ of $G$ is defined as the smallest height of a rooted tree that is valid for $G$.
Given positive integers $r$ and $c$, the graph of an $r \times c$ matrix $M$ is denoted by $G(M)$ and is constructed on the vertex set $[c]$. The pair $j,k$ is an edge if and only if there exists an $i \in [c]$ such that $M_{i,j} M_{i,k}\neq 0$. The \emph{dual tree-depth} of $M$ is defined as $td_G(M) := td(G(M^T))$. 

Notably, it has been proven that $td_G([A,B]^{(n)})\leq p+p'$ for any $n\in \N$ (see \cite[Lemma 96]{treedepth2022}). Using this bound with results in \cite[\S 7]{graver_complexity_onn2022}, it follows that $g(A,B)\leq (2\max(||A||_\infty, ||B||_\infty)+1)^{2^{(p+p')}-1}$. We would like to emphasize the significance of the block structure and sparsity within the design matrices when computing Graver bases. While these attributes have been utilized in optimization contexts (see \cite{N-fold2008}, \cite{StronglyPolyBlockStruct2018}, \cite{BlockStructILP2021}, \cite{treedepth2022}, \cite{ParamBlockStructIP2023}), their application in statistics remains relatively unexplored. 
\end{rmk}

In the upcoming section, we will delve into the scenario where a complete Markov basis is unavailable and explore the efforts made to address this challenge in specific contexts.

\subsection{When does an incomplete set of moves suffice?} \label{sec: simple moves}

One of the properties of a Markov basis $\mathcal M$ for a log-linear model with design matrix $A$ is that its elements generate the integer kernel of the matrix $A$, i.e., $\text{span}_\Z \mathcal M = \ker_\Z A$. Consequently, when a Markov basis is unavailable, a natural alternative is to explore ``simpler" subsets of $\ker_\Z A$ that can still generate the integer kernel, with the hope that they are sufficiently large to connect the fibers of the model. This has been a direction of research for some time with many open questions (see e.g., \cite{YoshidaOpenProblems2010}).
The most common subsets of the integer kernel used for this purpose are the following:

\begin{itemize}
	\item \emph{Lattice bases}. A set of vectors $\mathcal B$ in $\ker_\Z A$ is called a \emph{lattice basis} if it is linearly independent and $\text{span}_\Z \mathcal B =\ker_\Z A$.

	As noted in \cite[\S 1.3]{DSS09}, a lattice basis is a proper subset of a full Markov basis. While its size is fixed by the rank of $A$ and it is easy to compute, in general, it does not connect model fibers. The workaround  providing a provably connected chain is a simple idea: every Markov move is a linear combination of lattice basis moves. The difficulty is that how \emph{large} the linear combinations are required is not known.

	\item \emph{Circuits}. A vector $x\in \ker_\Z A$ is a \emph{circuit} if its support is minimal, i.e., if there is no vector $y\in \ker_\Z A$ such that $\text{supp}(y)\subset \text{supp}(x)$. The set of all such vectors is called the set of circuits of $A$. There are particular cases in which $A$ provides a nice combinatorial description for the set of circuits. Circuits are closely related to triangulations of the point configuration $A$.

\end{itemize}

In general one has the following containment relationships among the different subsets of $\ker_\Z A$: (1) every Markov basis contains a lattice basis (2) the Graver basis contains any minimal Markov basis and (3) the Graver basis contains the set of circuits. 

\begin{example} Let us revisit the $4\times 4$ independence model from Example~\ref{example: job data}, but we consider a synthetic data set described in \cite[p.371]{DS98}. Note that their discussion is about $10\times10$ tables, but for space considerations we consider a $4\times4$ sub-example, since it already proves our point. 
Namely, consider the following sparse table: 
\begin{table}[ht]
\centering
\begin{tabular}{r|rrrr}
  \hline
 & VeryD & LittleD & ModerateS & VeryS \\ 
  \hline
$<$ 15k & 1& 0& 0& 1\\ 
  15-25k & 1& 1& 0& 0\\ 
  25-40k & 0& 1& 1& 0\\ 
  $>$ 40k & 0& 0& 1& 1\\ 
   \hline
\end{tabular}
\end{table}

There are $282$ tables in the fiber of the model of independence corresponding to the margins of $v$. 
Using a lattice basis and a full Markov basis provides very different fiber simulations; see Figure~\ref{fig:Persi sparse table}. In particular, the Markov chain relying on lattice bases does not succeed to move at all, to any other table on the fiber,  in 5 out of the 10 simulated chains. 

\begin{figure}[h!]
	 \includegraphics[scale=.4]{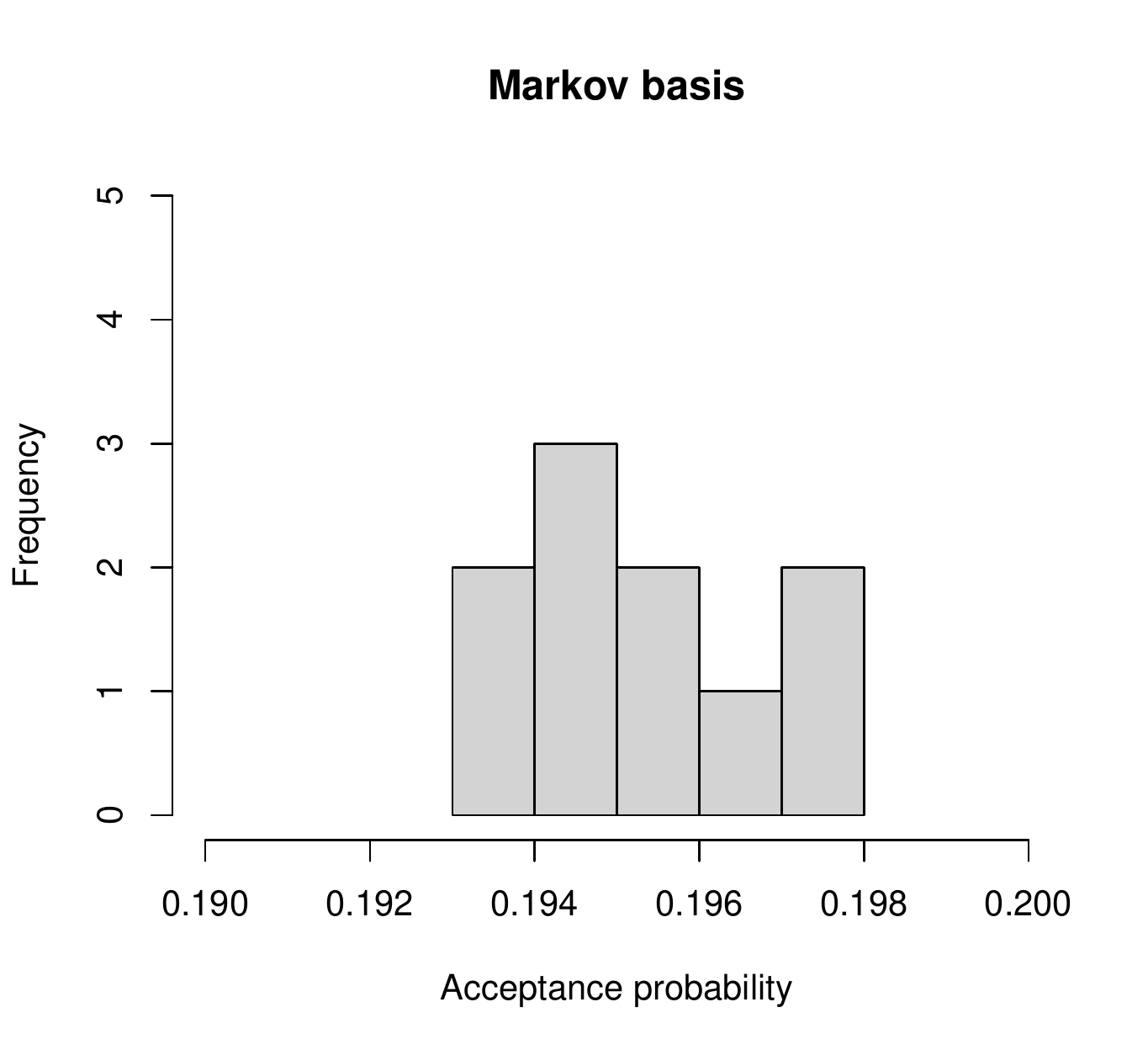}
 	 \includegraphics[scale=.4]{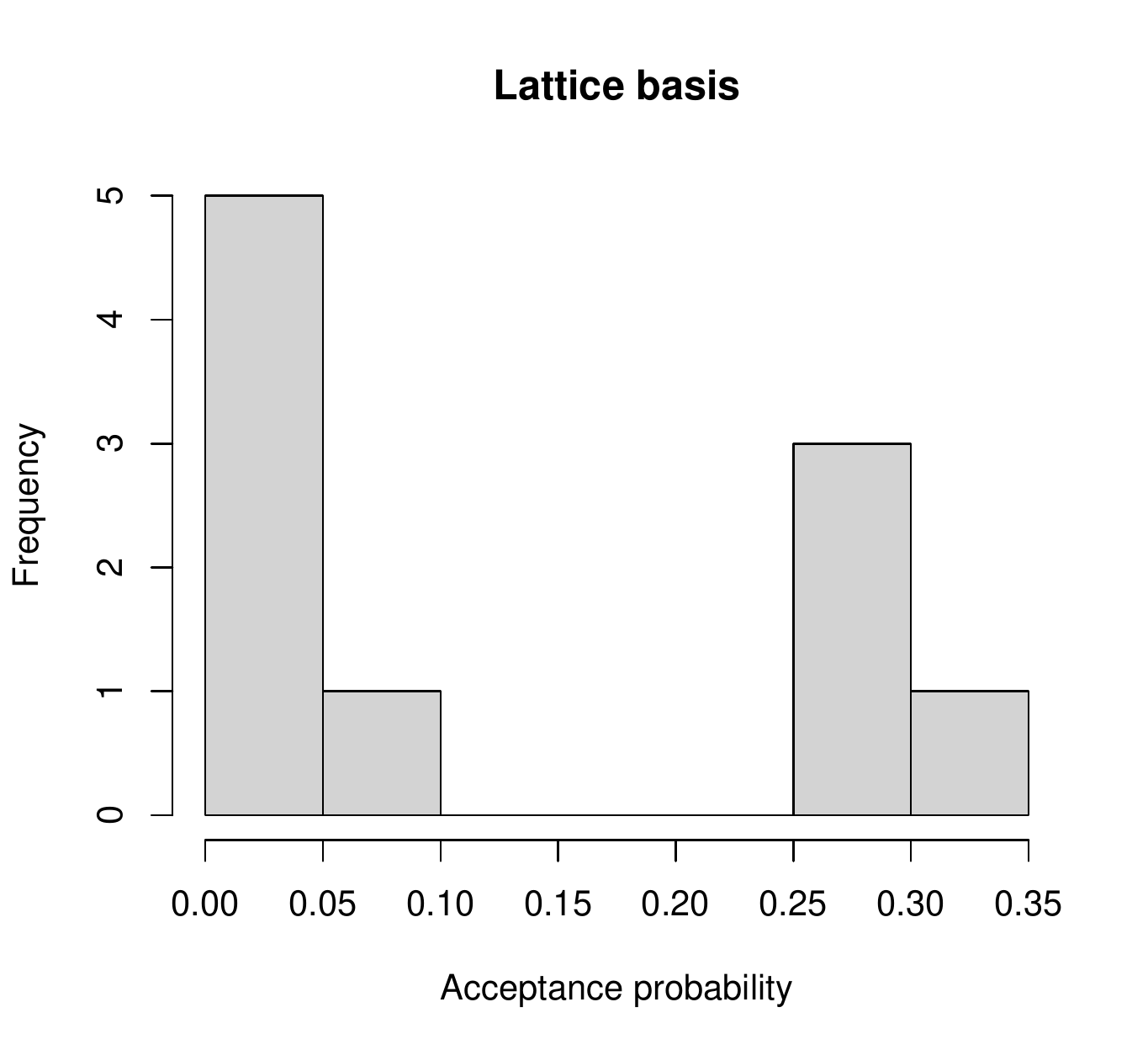}
\caption{Acceptance probabilities for proposed Markov moves for fiber samplers using two types of moves: (Left) a Markov basis and (Right) a lattice basis, for 10 repeated runs of a Markov chain of default length 10,000 using the R package {\tt algstat}.}\label{fig:Persi sparse table}
\end{figure} 

We note that the data does fit the independence model, and the Markov basis chain 
provides the $p$-value of $1$. 
To obtain another example with more interesting $p$-values and in which the lattice basis chain does explore the fiber more successfully, consider the following synthetic data table:

\begin{table}[ht]
\centering
\begin{tabular}{r|rrrr}
  \hline
 & VeryD & LittleD & ModerateS & VeryS \\ 
  \hline
$<$ 15k & 10& 0& 10& 0\\ 
  15-25k & 0& 3& 0& 3\\ 
  25-40k & 0& 0& 2& 40\\ 
  $>$ 40k & 2& 40& 0& 0\\ 
   \hline
\end{tabular}
\end{table}

For this example, there are $185,227,230$ tables in the fiber of the model of independence. 
Even if a lattice basis manages to explore the fiber (albeit at a slower rate, as the acceptance probability is usually lower than with Markov bases), the resulting $p$-values are not the same. Specifically, it appears that the lattice basis chain is missing a statistically significant part of the fiber, often not rejecting the model at the nominal level of $0.05$. 
\begin{figure}[h!]
	 \includegraphics[scale=.4]{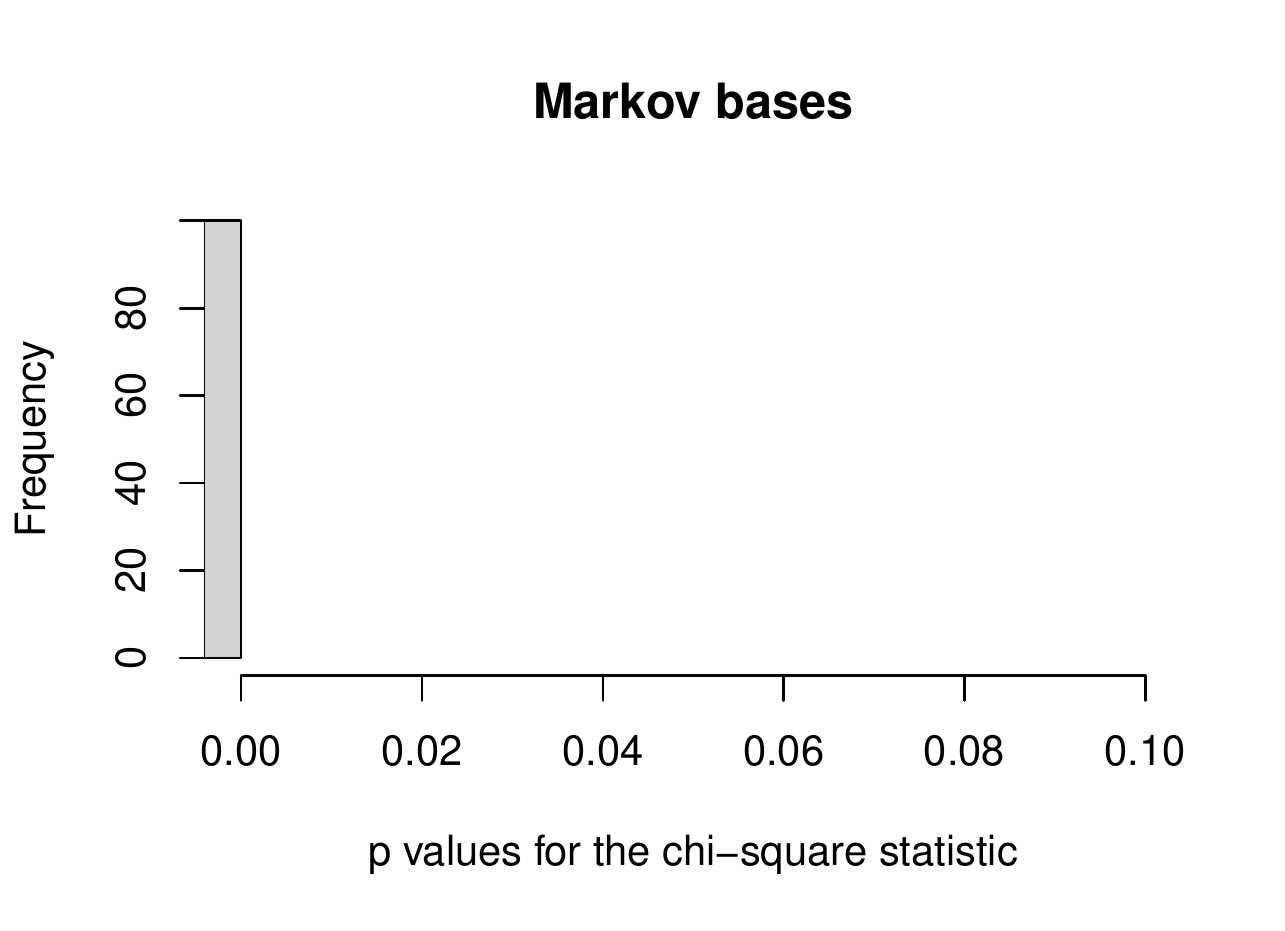}
 	 \includegraphics[scale=.4]{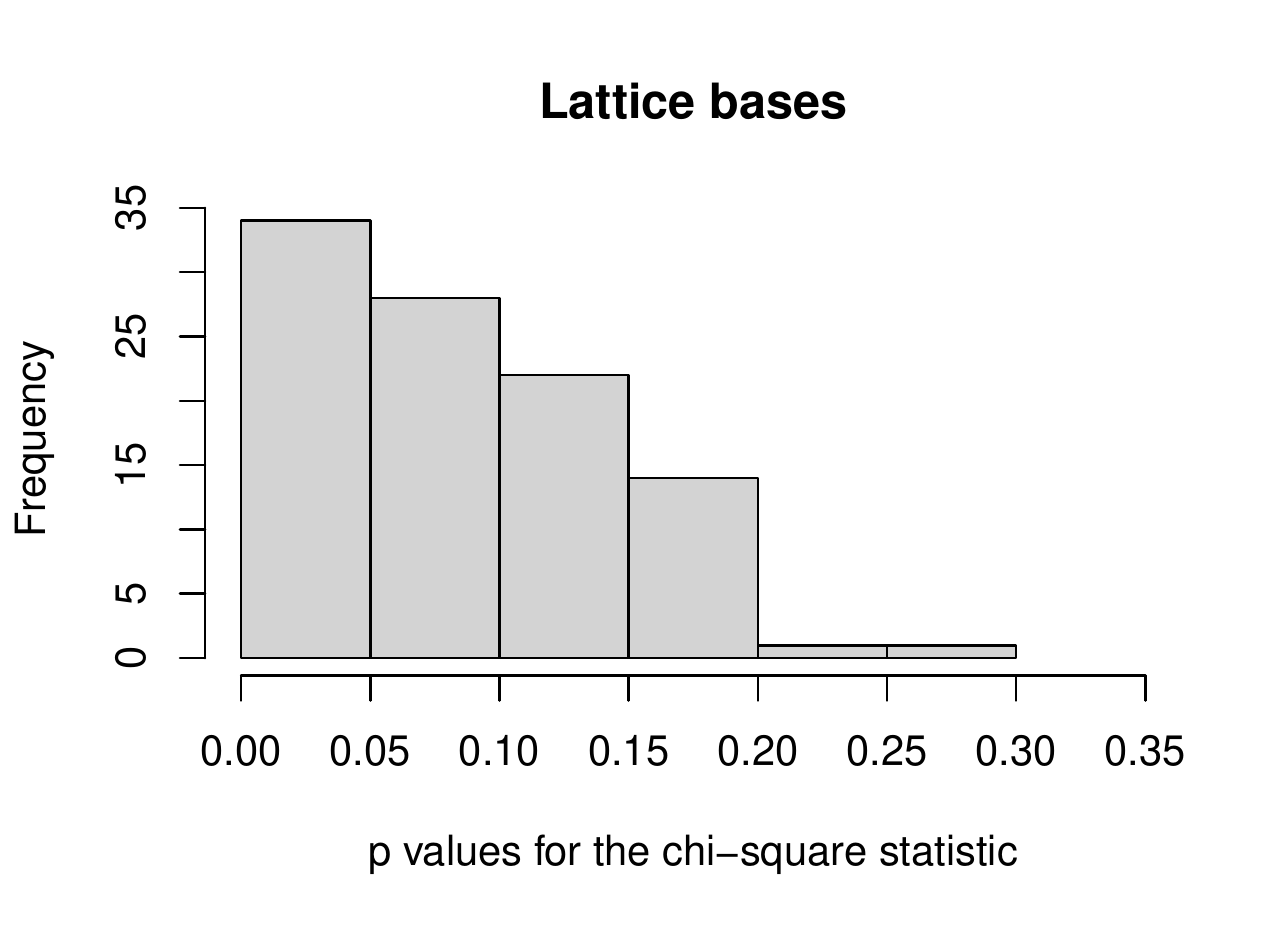}
\caption{Simulated $p$-values for the goodness-of-fit test of the independence model. The simulations are done using fiber samplers using two types of moves: (Left) a Markov basis and (Right) a lattice basis, for 100 repeated runs of a Markov chain of default length 10,000 using the R package {\tt algstat}.}\label{fig:Persi sparse table}
\end{figure} 
\end{example}

There are some special cases in which the algebra plays out very nicely and various bases of the model are equal, for example, when circuits are a Markov basis; see \cite[\S 3.2]{BesagClifford89} and \cite[\S 4.1]{geometricalMarkovChains}. In such cases, of course, circuits give an irreducible chain. Another is when a lattice basis is already a Markov basis. However, these scenarios are optimistic in the sense that such examples—say, models with unimodular design matrices—are extremely rare in statistical applications. In general, neither lattices basis nor circuits are Markov basis in which case we cannot blindly use them and expect to connect all the fibers of the model.

In \cite{geometricalMarkovChains} the authors examined a ``dynamic lattice basis" defined as the union of all lattice bases generated by reduced row echelon forms of the design matrix under column reordering. They proved that when $A$ is totally unimodular, the dynamic lattice basis is equal to the set of circuits of $A$, which, in turn, is equal to the Graver basis of $A$ according to \cite[Proposition 8.11]{St}. Since the Graver basis of a matrix contains any minimal Markov basis, it follows that the dynamic lattice basis is a Markov basis when $A$ is totally unimodular. However, unimodularity is a strong condition for a matrix to satisfy, as demonstrated by Seymour’s decomposition theorem (see \cite[\S 19.4]{Schrijver1999TheoryOL}). As an example of the study of unimodular matrices in specific models, we refer the reader to \cite{BersteinOneill-Hierarchical} and \cite{Berns-Sull-unimod17}, which provide a full description of hierarchical models with totally unimodular design matrices.

For the no-three-way interaction model many authors have considered a special set of moves, called \emph{basic moves}: elements of minimal $1$-norm which, like a lattice basis, span $\ker_\Z A$. The set of basic moves is not a Markov basis for the no-three-way interaction model on $I\times J\times K$ tables when at least two of $I, J, K$ are larger than 2 (see \cite[Chapter 9]{AHT2012}). However, this simple set of moves allows to connect some particular fibers in the model. \cite{BesagClifford89} stated  that ``basic moves  are irreducible in testing for the absence of three-way interaction in a $2\times J\times K$ contingency table whose layer totals are all  unity". 

\cite[Proposition 2]{BesagClifford89} also stated that basic moves suffice to connect $3$-way tables with \emph{positive margins}; a result that impacts \cite{RY10} who focused on bounded two-way contingency tables under independence model and show that in the absence of structural zeros, the set of basic moves suffice to connect the fiber. 

Another instance in which simple moves are enough to connect fibers with positive margins has been studied in \cite{KahleSullivantRauh}. In this work, the authors show that for certain graphical models, fibers with strictly positive margins can be connected using quadratic moves corresponding to conditional independent statements.

In the next section we will discuss some alternatives that have been used in order to connect fibers induced by an a priori incomplete set of moves.

\section{Relaxation of non-negativity of cells}\label{sec:relaxation}

Given the difficulties described in the previous section, what options remain at one's disposal? 
One approach that has shown some potential in practice is to use incomplete bases---easy to compute---in combination with another idea, namely relaxing the requirement that cells of the contingency table be non-negative. 
Of course, such tables with negative entries do not correspond to data observations, but one hopes that perhaps stepping temporarily outside of the `observable fiber' can lead one back to a connected chain. 
In this section, we review existing approaches based on the idea of relaxing cell entries to allow for negatives and formally define extended (to the negative) fibers, before proving that this approach will fail in general.

There are some special cases that work out quite nicely. 
For example, \cite[Proposition 3]{BuneaBesag} show that basic moves give an irreducible chain for the no-3-way interaction model on $2\times J\times K$ tables if one extends the fiber by allowing a single $-1$ entry in the table at any given step. Similarly, [\cite{Chen_etal-lattice-contingency}, Theorem 3.1] applied the non-negativity relaxation of the fibers to a logistic regression model, by allowing some entries to take $-1$ values. Motivated by these ideas, \cite{Lee2018} and \cite{Yoshida-Barnhill:Negative-cell} adopted the same approach to prove that basic moves induce an irreducible Markov chain on the fibers of the no-three-way interaction model on $3\times 3\times K$ and $3\times 4\times K$ tables while allowing temporary $-1$ entries. These findings made use of the full descriptions for the unique minimal Markov bases presented in \cite{Aoki-Takemura:3x3xK-tables} and \cite{Aoki-Takemura:3x4xK-tables}. In the upcoming subsections we formalize the non-negativity relaxation approach and study its limitations.  

\subsection{A general approach: non-negativity relaxation}

 The goal of the non-negativity relaxation approach will be to define irreducible Markov chains on the fibers $\fiberofb$ of a log-linear model defined by $A$ by using a simple set of moves and allowing ``temporary'' steps of the chain to be taken in relaxed fibers whose elements allow for negative values in some entries of the contingency tables. 

\begin{defn}\label{dfn:extended-fibers} 
Given an integer matrix $A$ as in \ref{defn:fiber} and a subset $S\subseteq [d_1]\times \cdots\times [d_k]$, we define for any $q\in \Z_{>0}$ the \emph{$(-q, S)$-relaxed fiber} of $b$ as 
\[
\mathcal F_{-q, S}(b):=\{u\in \Z^{d_1\times \cdots\times d_k}:Au=b \text{ and } u_{i_1, \ldots, i_k}\geq -q \mathds{1}_S(i_1, \ldots, i_k)\}.
\] 
where $\mathds{1}_S:[d_1]\times\cdots\times [d_k]\to \{0,1\}$ is the indicator function of $S$. We say that a set $\mathcal M\subset \ker_\Z A$  is a \emph{$(-q, S)$-Markov basis} for the model induced by $A$ if for every value of the marginal vector $b$ and any pair $u,v\in \fiberofb$ there is a choice of $L$ elements $m_1, \ldots, m_L\in \pm\mathcal{M}$ such that 
\[
u=v+\sum_{i=1}^L m_i \;\;\;\text{ and }\;\;\;v+\sum_{i=1}^\ell m_i\in \mathcal F_{-q, S}(b)\; \text{ for every }\;\ell\leq L.
\]

When $S=[d_1]\times\cdots\times [d_k]$ we write $\mathcal F_{-q}(b)$ instead of $\mathcal F_{-q, S}(b)$ and refer to this set as the $(-q)$-relaxed fiber of $b$. A $(-q, S)$-Markov basis in this case will be referred simply as a $(-q)$-Markov basis. 
\end{defn}

Notice that a $(-q, S)$-Markov basis of a matrix $A$ induces an irreducible Markov chain on every fiber as described in Algorithm \ref{alg:MCMC-alg}.

\begin{algorithm}[h]
\label{alg:MCMC-alg}
\SetKwInOut{Input}{Input}
\SetKwInOut{Output}{Output}
\Input{$u\in \mathcal F(b)$, a contingency table\\ 
      $\mathcal{M}$, a $(-q, S)$-Markov basis\\
      $\pi$, a desired distribution on $\mathcal F(b)$\\
      $N$, number of tables to compute}
\Output{A sequence of tables $\{u_n\}_{n=1}^N$ in $\mathcal F(b)$}
Set $u_1 \leftarrow u$\;

Set $v\leftarrow u$ (an auxiliary table)\;

\For {$n=1, \ldots, N$}{

	Choose $m \in \pm \mathcal{M}$ uniformly at random\;

	\eIf{$v+m\notin \mathcal{F}_{-q, S}(b)$}{
	  $u_{n+1} \leftarrow u_n$\;
	}{
	  \eIf{$v+m\in\mathcal F(b)$}{
	    $u_{n+1}\leftarrow v+m$ with probability $\min\left\{1, \frac{\pi(v+m)}{\pi(u_n)}\right\}$\;
	  }{

	    $u_{n+1}\leftarrow u_n$\;

	    $v \leftarrow v+m$\;
	    }
	  }
}

Return $\{u_n\}_{n=1}^N$

\caption{Metropolis-Hastings with a $(-q, S)$-Markov basis}
\end{algorithm}

The results mentioned in the opening paragraphs  of this section start out with a simple set of moves that are easy to construct, and then prove they are  a $(-1)$-Markov basis for their corresponding log-linear model. In general, a necessary condition for a ``simple'' set of moves $\mathcal M$ to be a $(-q)$-Markov basis is that it spans the integer kernel of the matrix $A$ with integer linear combinations. In such a case, Proposition \ref{prop:existence_q} guarantees the existence of a large enough relaxation on the fibers that would allow $\mathcal M$ to induce an irreducible Markov chain on the fibers as in \ref{alg:MCMC-alg}. The convergence of this chain to a desired distribution $\pi$ is ensured by moving from a state $u_n$ to a different state $u_{n+1}$ with acceptance probability $\min\big\{1, \frac{\pi(u_{n+1})}{\pi(u_n)}\big\}$.

\begin{proposition}[\cite{St}]\label{prop:existence_q} Let $\mathcal{M}\subset \ker_\Z A$ be such that $\text{span}_\Z(\mathcal M)=\ker_\Z A$. Then, there exists $q>0$ such that for every $u,v\in \fiberofb$ there is a choice of $L$ elements $m_1, \ldots, m_L\in \pm \mathcal M$ such that $u=v+\sum_{i=1}^L m_i$ with the requirement that 
\[
	v+\sum_{i=1}^\ell m_i\in \mathcal F_{-q}(b):=\Big\{u\in \Z^{d_1\cdots d_k}_{\geq -q}:Au=b\Big\}.
\]
In other words, we are able to connect the elements of $\fiberofb$ by allowing intermediate steps to be taken inside the extended fiber $\mathcal F_{-q}(b)$. 
\end{proposition}

Therefore, the general  {non-negativity relaxation approach} can be interpreted as follows.  

\begin{enumerate}
	\item Identify an easily attainable subset $\mathcal M\subset \ker_\Z A$ such that $\text{span}_\Z(\mathcal M)=\ker_\Z A$. As an example, we could compute a lattice basis $\mathcal M$ for $\ker_\Z A$.  

	\item Find a $q>0$ such that $\mathcal M$ is a $(-q, S)$-Markov basis for $A$ for some $S$.  
\end{enumerate}

Even though this strategy works in some special situations, it cannot be applied to every situation without a careful analysis of the connectivity on $\fiberofb$ using $\mathcal M$. In the absence of general bounds for $q$, it is necessary to prove irreducibility of the Markov chain on the fibers on a model-by-model case, based on the corresponding relaxation induced by a fixed value of $q$.  
Indeed, the following theorem shows that there is no global upper bound for $q$ in terms of the $1$-norm of $A$. 

\begin{thm}\label{thm:Unbounded_q} For any $N>0$, there exists a matrix $\Lambda_N$ with $||\Lambda_N||_1=4$ and $\mathcal M_N\subset \ker_\Z \Lambda_N$ such that $\text{span}_\Z(\mathcal M_N)=\ker_\Z \Lambda_N$ but $\mathcal M_N$ is not a $(-q)$-Markov basis for any $q<N$.  
\end{thm} 

The outcome of Theorem \ref{thm:Unbounded_q} is that having a set of moves that spans $\ker_\Z A$ is not enough to guarantee that a small relaxation of the fiber would be sufficient to create irreducible Markov chains on the fibers of the model. However, there are specific instances in the no-three-way interaction model where a simple set of moves has proven to be a $(-1)$-Markov basis when $I, J$ are fixed and $K$ grows, as previously mentioned. In the rest of the section, we aim to gain a better understanding of the limitations of the fiber relaxation technique when used on the no-three-way interaction model.

\subsection{Could relaxations work for the no-three-way interaction model?}\label{subsec:nothreeway}

A natural question arises: are the non-negativity constraints on the entries  responsible for the problematic behavior of the Markov basis for the model? In other words, we would like to investigate how large the elements of a $(-q,S)$-Markov basis can be when $I,J>0$ are unrestricted and $K=3$.

\medskip

For the remainder of our discussion, it is important to adopt the polyhedral viewpoint. Namely, that the fiber $\mathcal F(b)$ represents  integer points inside the polytope $P_{A,b}$, while $\mathcal F_{-q, S}(b)$ corresponds to the set of integer points of a new polytope $\widetilde P_{A,b}$. This polytope is defined as the space of solutions for the linear equations $Ax=b$ and is bounded by half-spaces $\widetilde H_{i_1, \ldots, i_k}:=\{x\in \R^{d_1\times\cdots\times d_k}:x_{i_1, \ldots, i_k}\geq -q\mathds{1}_S(i_1, \ldots, i_k)\}$, where we bound $\{x:Ax=b\}$ by translations of the hyperplanes that define the half-spaces $H_{i_1, \ldots, i_k}$.

The following theorem suggests that translating the non-negativity constraint hyperplanes by one unit can still lead to arbitrarily complicated elements inside any minimal Markov basis when $S$ is chosen poorly. The proof of this result is in the Appendix \ref{appendix:proofs}. 

\begin{thm}\label{thm:bad-news_restricted} For any nonnegative integer vector $\theta\in \N^\eta$, there are $I,J$ and $S\subset [I]\times [J]\times [3]$ with $|S|= 1+\sum_{i=1}^\eta\theta_i$, such that any minimal $(-1,S)$-Markov basis for the no-three-way interaction model on $I\times J\times 3$ tables must contain an element whose restriction to some $\eta$ entries is $\theta$ or $2\theta$.
\end{thm}  

Let us unpack the meaning and impact of Theorem~\ref{thm:bad-news_restricted}. We have seen in Corollary~\ref{cor:bad-news}  that Markov bases can be arbitrarily complicated for the no-3-way interaction model if two of the table dimensions are large. This results says that the same result holds for \emph{extended fibers} as well, when table cells are allowed to become negative. 
Indeed,  this means we have got bad news:  the existence of Markov bases elements as complicated as one wishes to see (or does not!) translates  even to the case of extended fibers! 

Given the universality result stated in Proposition~\ref{prop:all-linear-are-transportation}, the remainder of this section  focuses on the no-three-way interaction model. 
In particular, we study the effectiveness of non-negativity relaxation technique using basic moves, which for this model have received significant attention (see \cite{BesagClifford89}, \cite{Chen_etal-lattice-contingency}, \cite{HAT}). 

\begin{figure}[h!]
\centering
\includegraphics[width=0.35\textwidth]{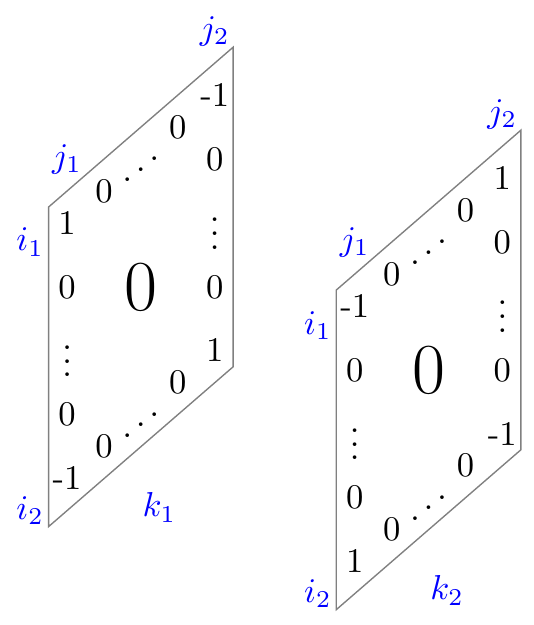}
\caption{The basic move $b(i_1, i_2; j_1, j_2; k_1, k_2)$}
\label{figure:basic-move}
\end{figure}

As mentioned in Section~\ref{sec: simple moves}, a \emph{basic move} for the no-three-way  model on $I\times J\times K$ tables is a zero-margin table with minimal $1$-norm. These basic moves can be described as tables $u=(u_{i,j,k})$ of the form
\[
u_{i,j,k}=
\begin{cases}
  1, & \text{if }(i,j,k)\in \{(i_1, j_1, k_1), (i_1, j_2, k_2), (i_2, j_1, k_2), (i_2, j_2, k_1)\}\\
  -1, & \text{if }(i,j,k)\in \{(i_2, j_2, k_2), (i_2, j_1, k_1), (i_1, j_2, k_1), (i_1, j_1, k_2)\}\\
  0, & \text{otherwise}
\end{cases}
\] 

\noindent for fixed indices $i_1\neq i_2\in [r]$, $j_1\neq j_2\in [c]$ and $k_1\neq k_2\in [3]$. We denote the basic move associated to these indices by $b (i_1,i_2;\; j_1, j_2;\; k_1, k_2)$ (see Figure \ref{figure:basic-move} for an illustration) and we denote the set of basic moves for the no-three-way interaction model on $I\times J\times K$ tables by $\mathcal{B}_{IJK}$ or simply by $\mathcal B$ when $I,J,K$ are clear from the context. 

It is known that for the design matrix $A$ of the no-three-way-interaction model for $I\times J\times K$ tables, any element in $\ker_\Z A$ can be written as a linear combination of the basic moves, i.e., $\text{span}_\Z (\mathcal B)=\ker_\Z A$ (see \cite{HAT}). Hence, Proposition \ref{prop:existence_q} guarantees the existence of a $q>0$ such that $\mathcal B$ is a $(-q)$-Markov basis for $A$. 

While it remains an open problem whether $\mathcal B_{IJK}$ is a $(-1)$-Markov basis for the no-three-way interaction model in general, it has been established, as mentioned earlier, that for specific cases such as $2\times J\times K$, $3 \times 3\times K$, and $ 4 \times 3\times K$ way tables, $\mathcal B$ is a $(-1)$-Markov basis for $A$. However, given the complex behavior of the fibers for $A$ described in Corollary~\ref{cor:bad-news}, it is hard to believe that the result generalizes when fixing $K=3$ and letting $I,J$ be unconstrained. To address this problem, we present the following partial result with proof in the Appendix \ref{appendix:proofs}.

\begin{proposition}\label{prop:anti-stair} Let $I,J\geq 3$ and let $S\subset [I]\times [J]\times [3]$ have an anti-staircase shape (defined in \ref{defn:anti-staircase}). Then, for any $q>0$ the set of basic moves is \emph{not} a $(-q, S)$-Markov basis for no-three-way interaction model on $I\times J \times 3$ tables.  
\end{proposition}

\begin{defn}\label{defn:anti-staircase} Let $I,J\geq 3$ and let $S\subset [I]\times [J]\times [3]$. We say that $S$ has a \textit{staircase shape} if there is a surjective function $\tau:[J]\to [3]$ or a surjective function $\tau':[I]\to [3]$ such that 
\[
S = \bigcup_{j=1}^J \{(i,j,\tau(j)): i \in [I]\} \;\;\;\text{ or }\;\;\; S = \bigcup_{i=1}^I \{(i,j,\tau'(i)): j \in [J]\}.
\] 
We say that $S$ has an \textit{anti-staircase shape} if $S$ is a complement of a subset of $[I]\times [J]\times[3]$ in staircase shape. As an example, the sets $S$ corresponding to the colored cells in Figure \ref{fig:anti-staircase} have a staircase shape.
\end{defn}

\newcommand{\firstcuboid}[4]{
\begin{tikzpicture}[scale=#4]


\fill[color=teal] (0,#2,0) -- (0,#2,#3) -- (#1/3,#2,#3) -- (#1/3,#2,0) -- cycle;

\fill[color=teal] (0,#2-1,#3) -- (0,#2,#3) -- (#1/3,#2,#3) -- (#1/3,#2-1,#3) -- cycle;
\fill[color=teal] (#1/3,#2-2,#3) -- (#1/3,#2-1,#3) -- (#1/3+3,#2-1,#3) -- (#1/3+3,#2-2,#3) -- cycle;
\fill[color=teal] (#1/3+3,#2-3,#3) -- (#1/3+3,#2-2,#3) -- (#1/3+4,#2-2,#3) -- (#1/3+4,#2-3,#3) -- cycle;

\fill[color=teal] (#1,0,0) -- (#1,0,#3) -- (#1,1,#3) -- (#1,1,0) -- cycle;


\foreach \x in {0,...,#1}
{ \draw[color=darkgray] (\x ,0 ,#3 ) -- (\x ,#2 ,#3 );
\draw[color=darkgray] (\x ,#2 ,#3 ) -- (\x ,#2 ,0 );
}

\foreach \y in {0,...,#2}
{ \draw[color=darkgray] (#1 ,\y ,#3 ) -- (#1 ,\y ,0 );
\draw[color=darkgray] (0 ,\y ,#3 ) -- (#1 ,\y ,#3 );
}

\foreach \z in {0,...,#3}
{ \draw[color=darkgray] (#1 ,0 ,\z ) -- (#1 ,#2 ,\z );
\draw[color=darkgray] (0 ,#2 ,\z ) -- (#1 ,#2 ,\z );
}

\end{tikzpicture}
}

\newcommand{\secondcuboid}[4]{
\begin{tikzpicture}[scale=#4]


\fill[color=teal] (0,#2,1) -- (0,#2,2) -- (#1,#2,2) -- (#1,#2,1) -- cycle;

\fill[color=teal] (0,0,#3) -- (0,1,#3) -- (#1,1,#3) -- (#1,0,#3) -- cycle;

\fill[color=teal] (#1,0,#3-2) -- (#1,0,#3) -- (#1,1,#3) -- (#1,1,#3-2) -- cycle;

\fill[color=teal] (#1,#2-1,1) -- (#1,#2-1,2) -- (#1,#2,2) -- (#1,#2,1) -- cycle;

\fill[color=teal] (#1,#2-2,0) -- (#1,#2-2,1) -- (#1,#2-1,1) -- (#1,#2-1,0) -- cycle;


\foreach \x in {0,...,#1}
{ \draw[color=darkgray] (\x ,0 ,#3 ) -- (\x ,#2 ,#3 );
\draw[color=darkgray] (\x ,#2 ,#3 ) -- (\x ,#2 ,0 );
}

\foreach \y in {0,...,#2}
{ \draw[color=darkgray] (#1 ,\y ,#3 ) -- (#1 ,\y ,0 );
\draw[color=darkgray] (0 ,\y ,#3 ) -- (#1 ,\y ,#3 );
}

\foreach \z in {0,...,#3}
{ \draw[color=darkgray] (#1 ,0 ,\z ) -- (#1 ,#2 ,\z );
\draw[color=darkgray] (0 ,#2 ,\z ) -- (#1 ,#2 ,\z );
}

\end{tikzpicture}
}

\newcommand{\thirdcuboid}[4]{
\begin{tikzpicture}[scale=#4]


\fill[color=teal] (2,#2,0) -- (2,#2,#3) -- (4,#2,#3) -- (4,#2,0) -- cycle;

\fill[color=teal] (0,#2-3,#3) -- (0,#2-2,#3) -- (1,#2-2,#3) -- (1,#2-3,#3) -- cycle;
\fill[color=teal] (1,#2-2,#3) -- (1,#2-1,#3) -- (2,#2-1,#3) -- (2,#2-2,#3) -- cycle;
\fill[color=teal] (4,#2-2,#3) -- (4,#2-1,#3) -- (5,#2-1,#3) -- (5,#2-2,#3) -- cycle;
\fill[color=teal] (2,#2-1,#3) -- (2,#2,#3) -- (4,#2,#3) -- (4,#2-1,#3) -- cycle;
\fill[color=teal] (4,#2-2,#3) -- (4,#2-1,#3) -- (5,#2-1,#3) -- (5,#2-2,#3) -- cycle;
\fill[color=teal] (#1/3+3,#2-3,#3) -- (#1/3+3,#2-2,#3) -- (#1/3+4,#2-2,#3) -- (#1/3+4,#2-3,#3) -- cycle;

\fill[color=teal] (#1,0,0) -- (#1,0,#3) -- (#1,1,#3) -- (#1,1,0) -- cycle;


\foreach \x in {0,...,#1}
{ \draw[color=darkgray] (\x ,0 ,#3 ) -- (\x ,#2 ,#3 );
\draw[color=darkgray] (\x ,#2 ,#3 ) -- (\x ,#2 ,0 );
}

\foreach \y in {0,...,#2}
{ \draw[color=darkgray] (#1 ,\y ,#3 ) -- (#1 ,\y ,0 );
\draw[color=darkgray] (0 ,\y ,#3 ) -- (#1 ,\y ,#3 );
}

\foreach \z in {0,...,#3}
{ \draw[color=darkgray] (#1 ,0 ,\z ) -- (#1 ,#2 ,\z );
\draw[color=darkgray] (0 ,#2 ,\z ) -- (#1 ,#2 ,\z );
}

\end{tikzpicture}
}

\begin{figure}[h]
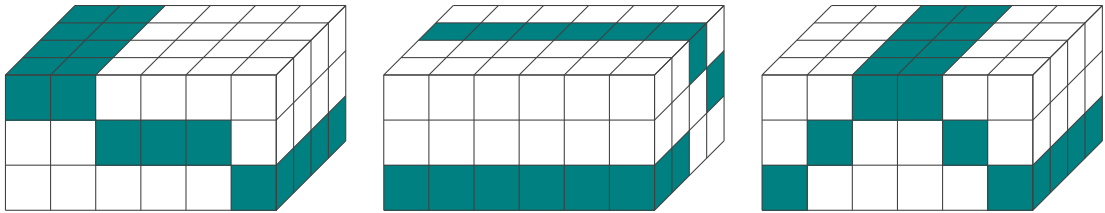
\label{fig:anti-staircase}
\centering
\begin{minipage}{0.3\textwidth}
  \firstcuboid{6}{3}{4}{.6}
\end{minipage}%
\begin{minipage}{0.3\textwidth}
  \secondcuboid{6}{3}{4}{.6}
\end{minipage}%
\begin{minipage}{0.3\textwidth}
  \thirdcuboid{6}{3}{4}{.6}
\end{minipage}  
\caption{Subsets of $[4]\times[6]\times[3]$ with staircase shape.}
\end{figure}

\section{Discussion} 

Proposition \ref{prop:all-linear-are-transportation} establishes the significance of the no-three-way interaction model for $I\times J\times 3$ tables in the realm of log-linear models. This is due to the fact that any fiber of a log-linear model is linearly isomorphic to a fiber of this particular model. Consequently, discovering a straightforward method to connect any fiber in the no-three-way interaction model would provide a means to define irreducible Markov chains on fibers within any log-linear model. As a result, the ability to address the following open questions becomes of utmost interest.

\begin{problem}\label{prob:basic-moves-connect}
Is the set of basic moves $\mathcal B$ a $(-1)$-Markov basis for the no-three-way interaction model on $I\times J\times 3$ tables for any $I,J\geq 3$?
\end{problem}

\begin{problem}\label{prob:finite-nonnegativity-relaxation}  Fix a positive integer $q>0$. Is it possible to give a general bound on the 1-norm of the elements of any minimal $(-q)$-Markov basis for the no-three-way interaction models on $I\times J\times 3$ tables?  
\end{problem}

In general, Markov complexity results can lead to tradeoffs in different directions. We have already seen that divide-and-conquer algorithms exist for decomposable hierarchical models. 
When the model is not decomposable, the bases might be too large to compute; mainstream 
 journals  have already published  works that point to various inefficacies of the method, e.g.,  \cite{DobraEtAl-IMA}, \cite{SteveAleMe-holland,PRF:09}. 

However, in these cases, one can resort to the so-called \emph{dynamic Markov bases}.  The idea behind dynamic Markov bases is simple, and can be described in two ways. The original one from \cite{Dob2012} is to compute a small set of moves that connects the local neighborhood of the fiber, and then expand on the fly, resulting in a fiber covering and therefore a connected chain.  Another one, with  proven success in network models \cite{GPS16,GPS21+,karwa2016exact,KP:AOAS}, is to use theoretical results about the \emph{structure} of the Markov basis moves, and then, instead of pre-computing the prohibitively large set, create an algorithm that creates one move at a time that can be applied to the current data point on the fiber. Importantly, many network models have 0/1 sampling constraints, and the bases are designed to work with those in mind, in line with  Theorem~\ref{thm:bounded graver is markov for bounded fibers}. See also \cite{RUMBA} for an algorithm that explores fibers using a combination of lattice or Markov moves and a biased iterative update to its sampling parameters. 

On the other hand, one could abandon pure MCMC methods entirely and devise alternative sampling algorithms. One such success story is \cite{KahleYoshidaGarciaPuente} who, motivated by the complexity of determining a Markov basis for some models, combine it with sequential importance sampling (SIS). Prior work on purely using SIS, however, was less impressive, as \cite{Dob2012} found, in numerical comparisons, that the dynamic Markov bases performed better than SIS approaches.

In terms of practical uses of Markov bases, at least two important ideas on speeding up Markov chains based on Markov bases deserve more attention and a more prominent place in the applied literature using Markov bases.  Namely, the idea of \emph{Hit and Run}  from  \cite{DiaconisAnderson} is applicable to \emph{any} Markov basis and makes a huge practical difference. The idea of blending Markov bases to create a faster, hybrid method was investigated in \cite{KahleYoshidaGarciaPuente}. Finally, there is the classical proposal by Besag, called ``the parallel method" \cite{BesagClifford89}, which results in an exchangeable sample from the conditional distribution.

We owe our readers a note about \emph{mixing times}. The study of mixing times of Markov chains arising from Markov bases is outside the scope of this paper, but the dynamic Markov bases references cited above show evidence---and in some cases, proof---of good or rapid mixing of the chains. These applied results stand in apparent  stark contrast to the theoretical advances in \cite{Tobias2015mixing}, who gives an asymptotic construction with fibers with low conductivity, and thus worst possible mixing time. In context of applications, this asymptotic result does not bring bad news to a specific data analysis instance, as it requires an asymptotically changing value of the sufficient statistic $t(U)$. Indeed, while \citeauthor{Tobias2015mixing} proves Markov bases do not suffice for good mixing behavior asymptotically, he does conclude that the chain has to be adapted for each specific fiber, that is, each specific value of $t(U)$. Therefore, the contrast of the theoretical vs. applied results indeed holds only in appearance.  See \cite{StanleyWindisch2018} for more on this topic.

While our paper focuses on the use of Markov bases for sampling from conditional distributions of log-affine models, and we avoid the use of algebra, the underlying algebraic connection has other uses in statistics.  Notably,  another line of work started in \cite{Diaconis-Baccalado-Holmes} relies on knowing a Markov basis of a discrete exponential family to extend de Finetti's idea to allow for a Bayesian analysis for almost exchangeable models. In \cite{Diaconis22approximateExchangability}, Markov bases are key to formulating partial exchangeability for contingency tables; see also \cite{Diaconis2022exchangabilityTables}. In another direction, \cite{karwa2016exact} extended the use of Markov bases and fiber sampling to \emph{mixtures} of log-linear models, specifically in the context of latent-variable network models. Finally, all of these methods focus on models on discrete random variables, but \cite{Diaconis-Holmes-Shahshahani} presented some theory, examples, and a set of open problems on continuous problems where one wants to do sampling from an exponential family conditional on sufficient statistics. This direction of investigation is yet to be fully explored.  \citeauthor{DiaconisEriksson2006}  investigate the Markov bases of the Birkhoff polytope (that is, the convex hull of the $n \times n$ permutation matrices). The associated statistical problem concerns the analysis of data where voters rank $n$ choices from most favored to least favored.

Finally, this survey is intended for an audience of statisticians, but the connection to algebra and combinatorics has inspired a wealth of non-statistical research, within that community. The Diaconis-Sturmfels theorem was very useful to combinatorial commutative algebraists because it gave a completely new way to think about the generators of ideals; the paper has over 800 citations to date, and a large number of those are theoretical developments. 
For our statistics audience, we note that the algebra and combinatorics literature contains hundreds of papers, but we wish to mention a few published papers that could be of interest for statisticians and point to other places where people have already computed Markov bases: In \cite{KahleSullivantRauh} the authors use an algebraic tool called \emph{binomial primary decomposition} 
to study the use of restricted sets of Markov moves, such as moves of degree two.  By investigating essential minimal generating elements of the Markov bases \cite{AokiTakemuraYoshida2008} showed how to recover some of the moves.
Similarly, in \cite{PetrovicStokes2013} 
the authors investigate Markov bases of hierarchical models. They give a universal construction of a class 
of minimal Markov bases in terms of the famous Stanley-Reisner rings. A series of papers generalizing the 
construction of the divide-and-conquer strategy for computing moves for decomposable models resulted in 
an algebraic operation, called a "toric fiber product", which then produced new algebraic insights and results; 
see, for example, \cite{Engstrom:2014aa},  \cite{RAUH2016276}. 
Combinatorialists have been attracted to investigate Markov bases due to their strong connections to graph theory.
The paper \cite{KNP10} and its references were concerned with $2 \times 2 \times \cdots \times 2$ tables, where all marginals are $2$-way marginals, the so called binary graph models. The Markov basis of the binary graph model of a $K_{4}$-minor free graph consists of binomials of degree $2$ and $4$ and they also show that the Markov basis of the binary graph model of $K_{n}$ requires binomials of degree $\Omega(n^{2- \epsilon})$. As we mentioned in Section 3, growth of Markov bases is a rather important topic and there has been vigorous activity, see 
\cite{Kosta2020}, \cite{TatakisThoma2022} and the references therein.

\section*{Acknowledgements} 

The authors are grateful to the anonymous referees whose comments and constructive feedback contributed to the improvement of this article. We thank Persi Diaconis and Thomas Kahle for reading a preliminary version of this article and providing further references and suggestions. We also thank Steffen Lauritzen for bringing Besag's 1989 parallel method to our attention several years ago.  We are grateful to Despina Stasi for early discussions and to Miles Bakenhus for his assistance running code with the LattE interface in \verb|R|. 
FAH and JDL are partially supported by NSF grant DMS-1818969. FAH is grateful for the support received through NSF TRIPODS Award no. CCF-1934568. SP is partially supported by DOE award \#1010629 and the Simons Foundation Collaboration Grant for Mathematicians \#854770. 

\bibliography{three-way-tables,AlgStatAndNtwksAndMB,StatsReferences}

\begin{thebibliography}{78}
\providecommand{\natexlab}[1]{#1}
\providecommand{\url}[1]{\texttt{#1}}
\expandafter\ifx\csname urlstyle\endcsname\relax
  \providecommand{\doi}[1]{doi: #1}\else
  \providecommand{\doi}{doi: \begingroup \urlstyle{rm}\Url}\fi

\bibitem[4ti2 team()]{4ti2}
4ti2 team.
\newblock 4ti2---a software package for algebraic, geometric and combinatorial
  problems on linear spaces.
\newblock Available at https://4ti2.github.io.

\bibitem[Agresti(2002)]{Agresti2002}
A.~Agresti.
\newblock \emph{Categorical Data Analysis}.
\newblock Wiley Series in Probability and Statistics. John Wiley \& Sons, Inc.,
  2002.

\bibitem[Andersen(1980)]{Andersen}
E.~B. Andersen.
\newblock \emph{Discrete Statistical Models with Social Science Applications}.
\newblock Elsevier, 1980.

\bibitem[Andersen and Diaconis(2007)]{DiaconisAnderson}
H.~C. Andersen and P.~Diaconis.
\newblock Hit and run as a unifying device.
\newblock \emph{Journal de la Soci\'et\'e fran\c{c}aise de statistique \& Revue
  de statistique appliqu\'ee}, 148\penalty0 (4):\penalty0 5--28, 2007.

\bibitem[Aoki and Takemura(2003{\natexlab{a}})]{Aoki-Takemura:3x3xK-tables}
S.~Aoki and A.~Takemura.
\newblock Minimal basis for a connected {M}arkov chain over $3 \times 3 \times
  k$ contingency tables with fixed two‐dimensional marginals.
\newblock \emph{Australian \& New Zealand Journal of Statistics}, 45,
  2003{\natexlab{a}}.

\bibitem[Aoki and Takemura(2003{\natexlab{b}})]{Aoki-Takemura:3x4xK-tables}
S.~Aoki and A.~Takemura.
\newblock The list of indispensable moves of the unique minimal {M}arkov basis
  for $3\times 4\times k$ and $4\times 4\times 4$ contingency tables with fixed
  two-dimensional marginal, 2003{\natexlab{b}}.

\bibitem[Aoki et~al.(2008)Aoki, Takemura, and Yoshida]{AokiTakemuraYoshida2008}
S.~Aoki, A.~Takemura, and R.~Yoshida.
\newblock Indispensable monomials of toric ideals and {M}arkov bases.
\newblock \emph{J. Symbolic Comput.}, 43\penalty0 (6-7):\penalty0 490--507,
  2008.
\newblock \doi{10.1016/j.jsc.2007.07.012}.

\bibitem[Aoki et~al.(2012)Aoki, Hara, and Takemura]{AHT2012}
S.~Aoki, H.~Hara, and A.~Takemura.
\newblock \emph{Markov Bases in Algebraic Statistics}.
\newblock Springer Series in Statistics. Springer New York, 2012.

\bibitem[Bacallado et~al.(2015)Bacallado, Diaconis, and
  Holmes]{Diaconis-Baccalado-Holmes}
S.~Bacallado, P.~Diaconis, and S.~Holmes.
\newblock de {F}inetti priors using {M}arkov chain {M}onte {C}arlo
  computations.
\newblock \emph{Statistics and computing}, 25\penalty0 (4):\penalty0 797--808,
  2015.

\bibitem[Bakenhus and Petrovi\'c(To appear)]{RUMBA}
M.~Bakenhus and S.~Petrovi\'c.
\newblock Sampling lattice points in a polytope: a bayesian biased algorithm
  with random updates.
\newblock \emph{Algebraic Statistics}, To appear.

\bibitem[Barndorff-Nielsen(1978)]{Barndorff-Nielsen}
O.~Barndorff-Nielsen.
\newblock \emph{Information and Exponential Families: In Statistical Theory},
  volume Reprinted in 2014.
\newblock Wiley Series in Probability and Statistics, 1978.

\bibitem[Bernstein and O'Neill(2017)]{BersteinOneill-Hierarchical}
D.~Bernstein and C.~O'Neill.
\newblock Unimodular hierarchical models and their {G}raver bases.
\newblock \emph{Journal of Algebraic Statistics}, 8, 04 2017.

\bibitem[Bernstein and Sullivant(2017)]{Berns-Sull-unimod17}
D.~Bernstein and S.~Sullivant.
\newblock Unimodular binary hierarchical models.
\newblock \emph{Journal of Combinatorial Theory, Series B}, 123:\penalty0
  97--125, 2017.

\bibitem[Berstein and Onn(2009)]{graver_complexity_onn2009}
Y.~Berstein and S.~Onn.
\newblock The {G}raver complexity of integer programming.
\newblock \emph{Annals of Combinatorics}, 13:\penalty0 289--296, 11 2009.

\bibitem[Besag and Clifford(1989)]{BesagClifford89}
J.~Besag and P.~Clifford.
\newblock Generalized {M}onte {C}arlo significance tests.
\newblock \emph{Biometrika}, 76\penalty0 (4):\penalty0 633--642, 1989.

\bibitem[Besag and Mondal(2013)]{BesagMondal}
J.~Besag and D.~Mondal.
\newblock Exact goodness-of-fit tests for {M}arkov chains.
\newblock \emph{Biometrics}, 69\penalty0 (2), 2013.

\bibitem[Bunea and Besag(2000)]{BuneaBesag}
F.~Bunea and J.~Besag.
\newblock {MCMC} in $i\times j\times k$ contingency tables.
\newblock \emph{Fields Institute Communications}, 26, 2000.

\bibitem[Campo et~al.(2017)Campo, Cepeda, and Uhler]{DelCampoCepedaUhler2017}
A.~M.~D. Campo, S.~Cepeda, and C.~Uhler.
\newblock Exact goodness-of-fit testing for the {I}sing model.
\newblock \emph{Scandinavian Journal of Statistics}, 44\penalty0 (2):\penalty0
  285--306, 2017.

\bibitem[Chen(2007)]{Chen-2way-zeros}
Y.~Chen.
\newblock Conditional inference on tables with structural zeros.
\newblock \emph{Journal of Computational and Graphical Statistics}, 16\penalty0
  (2), 2007.

\bibitem[Chen et~al.(2005)Chen, Dinwoodie, and
  Dobra]{Chen_etal-lattice-contingency}
Y.~Chen, I.~Dinwoodie, and A.~Dobra.
\newblock Lattice points, contingency tables, and sampling.
\newblock \emph{Contemporary Mathematics}, 374, 2005.

\bibitem[Cox et~al.(2015)Cox, Little, and O'Shea]{CLO}
D.~A. Cox, J.~B. Little, and D.~O'Shea.
\newblock \emph{Ideals, Varieties, and Algorithms}.
\newblock Springer, 4th edition, 2015.

\bibitem[Cslovjecsek et~al.(2021)Cslovjecsek, Eisenbrand, Hunkenschr\"{o}der,
  Rohwedder, and Weismantel]{BlockStructILP2021}
J.~Cslovjecsek, F.~Eisenbrand, C.~Hunkenschr\"{o}der, L.~Rohwedder, and
  R.~Weismantel.
\newblock Block-structured integer and linear programming in strongly
  polynomial and near linear time.
\newblock In \emph{Proceedings of the Thirty-Second Annual ACM-SIAM Symposium
  on Discrete Algorithms}, SODA '21, page 1666–1681. Society for Industrial
  and Applied Mathematics, 2021.

\bibitem[Cslovjecsek et~al.(2023)Cslovjecsek, Koutecký, Lassota, Pilipczuk,
  and Polak]{ParamBlockStructIP2023}
J.~Cslovjecsek, M.~Koutecký, A.~Lassota, M.~Pilipczuk, and A.~Polak.
\newblock Parameterized algorithms for block-structured integer programs with
  large entries, 2023.
\newblock To appear in SODA 2024. Available at
  \href{https://arxiv.org/abs/2311.01890}{arXiv:2311.01890}.

\bibitem[De~Loera and Onn(2006{\natexlab{a}})]{slimTables}
J.~De~Loera and S.~Onn.
\newblock Markov bases of three-way tables are arbitrarily complicated.
\newblock \emph{J. Symbolic Comput.}, 41\penalty0 (2):\penalty0 173--181,
  2006{\natexlab{a}}.

\bibitem[De~Loera and Onn(2006{\natexlab{b}})]{DeLoera-Onn_all_linear}
J.~A. De~Loera and S.~Onn.
\newblock All linear and integer programs are slim 3‐way transportation
  programs.
\newblock \emph{SIAM Journal on Optimization}, 17\penalty0 (3):\penalty0
  806--821, 2006{\natexlab{b}}.

\bibitem[{De Loera} et~al.(2008){De Loera}, Hemmecke, Onn, and
  Weismantel]{N-fold2008}
J.~A. {De Loera}, R.~Hemmecke, S.~Onn, and R.~Weismantel.
\newblock N-fold integer programming.
\newblock \emph{Discrete Optimization}, 5\penalty0 (2):\penalty0 231--241,
  2008.

\bibitem[Diaconis(2022{\natexlab{a}})]{Diaconis2022exchangabilityTables}
P.~Diaconis.
\newblock Partial exchangeability for contingency tables.
\newblock \emph{Mathematics}, 10\penalty0 (3):\penalty0 442,
  2022{\natexlab{a}}.

\bibitem[Diaconis(2022{\natexlab{b}})]{Diaconis22approximateExchangability}
P.~Diaconis.
\newblock Approximate exchangeability and de finetti priors in 2022.
\newblock \emph{Scandinavian Journal of Statistics}, 50\penalty0 (1):\penalty0
  38--53, 2022{\natexlab{b}}.

\bibitem[Diaconis and Eriksson(2006)]{DiaconisEriksson2006}
P.~Diaconis and N.~Eriksson.
\newblock Markov bases for noncommutative {F}ourier analysis of ranked data.
\newblock \emph{J. Symbolic Comput.}, 41\penalty0 (2):\penalty0 182--195, 2006.
\newblock \doi{10.1016/j.jsc.2005.04.009}.

\bibitem[Diaconis and Sturmfels(1998)]{DS98}
P.~Diaconis and B.~Sturmfels.
\newblock Algebraic algorithms for sampling from conditional distributions.
\newblock \emph{Annals of Statistics}, 26\penalty0 (1):\penalty0 363--397,
  1998.

\bibitem[Diaconis et~al.(2013)Diaconis, Holmes, and
  Shahshahani]{Diaconis-Holmes-Shahshahani}
P.~Diaconis, S.~Holmes, and M.~Shahshahani.
\newblock Sampling from a manifold.
\newblock \emph{Institute of Mathematical Statistics}, pages 102--125, 2013.

\bibitem[Dobra(2003)]{Dob03}
A.~Dobra.
\newblock Markov bases for decomposable graphical models.
\newblock \emph{Bernoulli}, 9\penalty0 (6):\penalty0 1093--1108, 2003.

\bibitem[Dobra(2012)]{Dob2012}
A.~Dobra.
\newblock Dynamic {M}arkov bases.
\newblock \emph{Journal of Computational and Graphical Statistics}, pages
  496--517, 2012.

\bibitem[Dobra and Sullivant(2004)]{DS04}
A.~Dobra and S.~Sullivant.
\newblock A divide-and-conquer algorithm for generating {M}arkov bases of
  multi-way tables.
\newblock \emph{Computational Statistics}, 19:\penalty0 347--366, 2004.

\bibitem[Dobra et~al.(2008)Dobra, Fienberg, Rinaldo, Slavkovi\'c, and
  Zhou]{DobraEtAl-IMA}
A.~Dobra, S.~E. Fienberg, A.~Rinaldo, A.~Slavkovi\'c, and Y.~Zhou.
\newblock Algebraic statistics and contingency table problems: Log-linear
  models, likelihood estimation and disclosure limitation.
\newblock In \emph{In IMA Volumes in Mathematics and its Applications: Emerging
  Applications of Algebraic Geometry}, pages 63--88. Springer Science+Business
  Media, Inc, 2008.

\bibitem[Drton et~al.(2009)Drton, Sturmfels, and Sullivant]{DSS09}
M.~Drton, B.~Sturmfels, and S.~Sullivant.
\newblock \emph{Lectures on Algebraic Statistics}, volume~39 of
  \emph{Oberwolfach Seminars}.
\newblock Birkh{\"{a}}user, 2009.

\bibitem[Eisenbrand et~al.(2022)Eisenbrand, Hunkenschr{\"o}der, Klein,
  Kouteck{\'y}, Levin, and Onn]{treedepth2022}
F.~Eisenbrand, C.~Hunkenschr{\"o}der, K.-M. Klein, M.~Kouteck{\'y}, A.~Levin,
  and S.~Onn.
\newblock An algorithmic theory of integer programming, 2022.
\newblock Available at
  \href{https://arxiv.org/abs/1904.01361}{arXiv:1904.01361}.

\bibitem[Engstr{\"o}m et~al.(2014)Engstr{\"o}m, Kahle, and
  Sullivant]{Engstrom:2014aa}
A.~Engstr{\"o}m, T.~Kahle, and S.~Sullivant.
\newblock Multigraded commutative algebra of graph decompositions.
\newblock \emph{Journal of Algebraic Combinatorics}, 39\penalty0 (2):\penalty0
  335--372, 2014.
\newblock \doi{10.1007/s10801-013-0450-0}.

\bibitem[Fienberg(1980)]{CategoricalDataAnalysis}
S.~E. Fienberg.
\newblock \emph{The Analysis of Cross-Classified Categorical Data}.
\newblock Springer, Reprinted 2007, 1980.

\bibitem[Fienberg and Wasserman(1981)]{FienbergWasserman1981categorical}
S.~E. Fienberg and S.~S. Wasserman.
\newblock Categorical data analysis of single sociometric relations.
\newblock \emph{Sociological methodology}, 12:\penalty0 156--192, 1981.

\bibitem[Fienberg et~al.(2010)Fienberg, Petrovi\'c, and
  Rinaldo]{SteveAleMe-holland}
S.~E. Fienberg, S.~Petrovi\'c, and A.~Rinaldo.
\newblock \emph{Algebraic statistics for $p_1$ random graph models: Markov
  bases and their uses}, volume Papers in Honor of Paul W. Holland, ETS.
\newblock Springer, 2010.

\bibitem[Geiger et~al.(2006)Geiger, Meek, and Sturmfels]{GeigerMeekSturmfels}
D.~Geiger, C.~Meek, and B.~Sturmfels.
\newblock On the toric algebra of graphical models.
\newblock \emph{Annals of Statistics}, 34:\penalty0 1463--1492, 2006.

\bibitem[Goldenberg et~al.(2010)Goldenberg, Zheng, Fienberg, and
  Airoldi]{goldenberg2010survey}
A.~Goldenberg, A.~X. Zheng, S.~E. Fienberg, and E.~M. Airoldi.
\newblock A survey of statistical network models.
\newblock \emph{Foundations and Trends{\textregistered} in Machine Learning},
  2\penalty0 (2):\penalty0 129--233, 2010.

\bibitem[Gross et~al.(2016)Gross, Petrovi\'c, and Stasi]{GPS16}
E.~Gross, S.~Petrovi\'c, and D.~Stasi.
\newblock {G}oodness-of-fit for log-linear network models: {D}ynamic {M}arkov
  bases using hypergraphs.
\newblock \emph{Annals of the Institute of Statistical Mathematics}, 2016.

\bibitem[Gross et~al.(2021)Gross, Petrovi\'c, and Stasi]{GPS21+}
E.~Gross, S.~Petrovi\'c, and D.~Stasi.
\newblock Goodness of fit for log-linear {ERGM}s.
\newblock Preprint. In revision, 2021.

\bibitem[Hara and Takemura(2010)]{HT:10}
H.~Hara and A.~Takemura.
\newblock Connecting tables with zero-one entries by a subset of a markov
  basis.
\newblock In M.~Viana and H.~Wynn, editors, \emph{Algebraic Methods in
  Statistics and Probability II}, volume 516 of \emph{Contemporary
  Mathematics}. American Mathematical Society, 2010.

\bibitem[Hara et~al.(2012)Hara, Aoki, and Takemura]{HAT}
H.~Hara, S.~Aoki, and A.~Takemura.
\newblock Running {M}arkov chain without {M}arkov basis.
\newblock In \emph{Harmony of Gr{\"o}bner Bases and the Modern Industrial
  Society}, pages 46--62. World Scientific, 2012.

\bibitem[Hazelton et~al.(2020)Hazelton, Mcveagh, and van
  Brunt]{geometricalMarkovChains}
M.~L. Hazelton, M.~R. Mcveagh, and B.~van Brunt.
\newblock {Geometrically aware dynamic Markov bases for statistical linear
  inverse problems}.
\newblock \emph{Biometrika}, 108\penalty0 (3):\penalty0 609--626, 2020.

\bibitem[Ho{\c{s}}ten and Sullivant(2007)]{HoSu}
S.~Ho{\c{s}}ten and S.~Sullivant.
\newblock A finiteness theorem for {M}arkov bases of hierarchical models.
\newblock \emph{J. Combin. Theory Ser. A}, 114\penalty0 (2):\penalty0 311--321,
  2007.

\bibitem[Kahle et~al.(2014{\natexlab{a}})Kahle, Garcia-Puente, and
  Yoshida]{algstat.R}
D.~Kahle, L.~Garcia-Puente, and R.~Yoshida.
\newblock algstat: Algebraic statistics in {R}.
\newblock R package version 0.1.1, url=\url{https://github.com/dkahle/algstat},
  2014{\natexlab{a}}.

\bibitem[Kahle et~al.(2018)Kahle, Yoshida, and
  Garcia-Puente]{KahleYoshidaGarciaPuente}
D.~Kahle, R.~Yoshida, and L.~Garcia-Puente.
\newblock Hybrid schemes for exact conditional inference in discrete
  exponential families.
\newblock \emph{Annals of the Institute of Statistical Mathematics},
  70\penalty0 (5):\penalty0 983--1011, 2018.

\bibitem[Kahle et~al.(2014{\natexlab{b}})Kahle, Rauh, and
  Sullivant]{KahleSullivantRauh}
T.~Kahle, J.~Rauh, and S.~Sullivant.
\newblock {Positive margins and primary decomposition}.
\newblock \emph{Journal of Commutative Algebra}, 6\penalty0 (2):\penalty0 173
  -- 208, 2014{\natexlab{b}}.
\newblock \doi{10.1216/JCA-2014-6-2-173}.

\bibitem[Karwa and Petrovi\'c(2016)]{KP:AOAS}
V.~Karwa and S.~Petrovi\'c.
\newblock Coauthorship and citation networks for statisticians: Comment.
  invited comment on the paper by {J}in and {J}i.
\newblock \emph{Annals of Applied Statistics}, 10\penalty0 (4):\penalty0
  1827--1834, 2016.

\bibitem[Karwa et~al.(2023)Karwa, Pati, Petrovi{\'c}, Solus, Alexeev,
  Rai{\v{c}}, Wilburne, Williams, and Yan]{karwa2016exact}
V.~Karwa, D.~Pati, S.~Petrovi{\'c}, L.~Solus, N.~Alexeev, M.~Rai{\v{c}},
  D.~Wilburne, R.~Williams, and B.~Yan.
\newblock Monte {C}arlo goodness-of-fit tests for degree corrected and related
  stochastic blockmodels.
\newblock \emph{Journal of the Royal Statistical Society, series B}, To appear,
  2023.

\bibitem[Knop et~al.(2020)Knop, Pilipczuk, and Wrochna]{Knop_td_bound2020}
D.~Knop, M.~Pilipczuk, and M.~Wrochna.
\newblock Tight complexity lower bounds for integer linear programming with few
  constraints.
\newblock \emph{ACM Trans. Comput. Theory}, 12\penalty0 (3), 2020.

\bibitem[Kosta(2020)]{Kosta2020}
D.~Kosta.
\newblock Markov bases of toric ideals: connecting commutative algebra and
  statistics.
\newblock \emph{Lond. Math. Soc. Newsl.}, \penalty0 (486):\penalty0 34--37,
  2020.
\newblock ISSN 2516-3841.

\bibitem[Kouteck\'{y} et~al.(2018)Kouteck\'{y}, Levin, and
  Onn]{StronglyPolyBlockStruct2018}
M.~Kouteck\'{y}, A.~Levin, and S.~Onn.
\newblock {A Parameterized Strongly Polynomial Algorithm for Block Structured
  Integer Programs}.
\newblock In \emph{45th International Colloquium on Automata, Languages, and
  Programming (ICALP 2018)}, volume 107 of \emph{Leibniz International
  Proceedings in Informatics (LIPIcs)}, pages 85:1--85:14, 2018.

\bibitem[Kr\`al et~al.(2010)Kr\`al, Norine, and Pangra\`c]{KNP10}
D.~Kr\`al, S.~Norine, and O.~Pangra\`c.
\newblock Markov bases of binary graph models of {$K_4$}-minor free graphs.
\newblock \emph{Journal of Combinatorial Theory, Series A}, 117\penalty0
  (6):\penalty0 759--765, 2010.

\bibitem[Lauritzen(1996)]{Lauritzen}
S.~L. Lauritzen.
\newblock \emph{Graphical models}.
\newblock Oxford Statistical Science Series, 1996.

\bibitem[Lee(2018)]{Lee2018}
S.~Lee.
\newblock Markov chain {M}onte {C}arlo and exact conditional tests with
  three-way contingency tables.
\newblock Master's thesis, Naval Postgraduate School, 2018.

\bibitem[Ogawa et~al.(2013)Ogawa, Hara, and Takemura]{OHT:11}
M.~Ogawa, H.~Hara, and A.~Takemura.
\newblock Graver basis for an undirected graph and its application to testing
  the beta model of random graphs.
\newblock \emph{Annals of Institute of Statistical Mathematics.}, 65:\penalty0
  191--212, 2013.

\bibitem[Onn et~al.(2022)Onn, Thoma, and Vladoiu]{graver_complexity_onn2022}
S.~Onn, A.~Thoma, and M.~Vladoiu.
\newblock Asymptotic behavior of {M}arkov complexity of matrices, 2022.

\bibitem[Petrovi\'{c} and Stokes(2013)]{PetrovicStokes2013}
S.~Petrovi\'{c} and E.~Stokes.
\newblock Betti numbers of {S}tanley-{R}eisner rings determine hierarchical
  {M}arkov degrees.
\newblock \emph{J. Algebraic Combin.}, 37\penalty0 (4):\penalty0 667--682,
  2013.
\newblock \doi{10.1007/s10801-012-0381-1}.

\bibitem[Petrovi{\'c} et~al.(2010)Petrovi{\'c}, Rinaldo, and Fienberg]{PRF:09}
S.~Petrovi{\'c}, A.~Rinaldo, and S.~E. Fienberg.
\newblock Algebraic statistics for a directed random graph model with
  reciprocation.
\newblock In M.~Viana and H.~Wynn, editors, \emph{Algebraic Methods in
  Statistics and Probability II}, volume 516 of \emph{Contemporary
  Mathematics}, pages 261--283. American Mathematical Society, Providence RI,
  2010.

\bibitem[Rapallo and Yoshida(2010)]{RY10}
F.~Rapallo and R.~Yoshida.
\newblock Markov bases and subbases for bounded contingency tables.
\newblock \emph{Annals of the Institute of Statistical Mathematics},
  62\penalty0 (4):\penalty0 785--805, 2010.

\bibitem[Rasch(1960)]{Rasch}
G.~Rasch.
\newblock \emph{Probabilistic Models for Some Intelligence and Attainment
  Tests}.
\newblock University of Chicago Press, Chicago., 1960.

\bibitem[Rauh and Sullivant(2016)]{RAUH2016276}
J.~Rauh and S.~Sullivant.
\newblock Lifting markov bases and higher codimension toric fiber products.
\newblock \emph{Journal of Symbolic Computation}, 74:\penalty0 276--307, 2016.
\newblock \doi{https://doi.org/10.1016/j.jsc.2015.07.003}.

\bibitem[Rinaldo et~al.(2013)Rinaldo, Petrovi\'c, and Fienberg]{RPF:11}
A.~Rinaldo, S.~Petrovi\'c, and S.~E. Fienberg.
\newblock Maximum likelihood estimation in the {B}eta model.
\newblock \emph{Annals of Statistics}, 41\penalty0 (3):\penalty0 1085--1110,
  2013.

\bibitem[Schrijver(1999)]{Schrijver1999TheoryOL}
A.~Schrijver.
\newblock Theory of linear and integer programming.
\newblock In \emph{Wiley-Interscience series in discrete mathematics and
  optimization}, 1999.

\bibitem[Stanley and Windisch(2018)]{StanleyWindisch2018}
C.~Stanley and T.~Windisch.
\newblock Heat-bath random walks with markov bases.
\newblock \emph{Advances in Applied Mathematics}, 92:\penalty0 122--143, 2018.

\bibitem[Sturmfels(1996)]{St}
B.~Sturmfels.
\newblock \emph{Gr{\"{o}}bner bases and convex polytopes}.
\newblock University Lecture Series, no. 8, American Mathematical Society,
  1996.

\bibitem[Sullivant(2021)]{SethBook}
S.~Sullivant.
\newblock \emph{Algebraic Statistics}.
\newblock Graduate Studies in Mathematics. American Mathematical Society, 2021.

\bibitem[Takken(2000)]{takken2000}
A.~Takken.
\newblock \emph{Monte Carlo Goodness-of-Fit Tests for Discrete Data}.
\newblock PhD thesis, Dept. Statistics, Stanford University, 2000.

\bibitem[Tatakis and Thoma(2022)]{TatakisThoma2022}
C.~Tatakis and A.~Thoma.
\newblock On the relative size of toric bases.
\newblock \emph{Journal of Algebra and Its Applications}, 21\penalty0
  (04):\penalty0 2250079, 2022.

\bibitem[Windisch(2015)]{Tobias2015mixing}
T.~Windisch.
\newblock Rapid mixing and {M}arkov bases.
\newblock \emph{{SIAM} journal of discrete mathematics}, 30\penalty0 (4), 2015.
\newblock Available at
  \href{https://arxiv.org/abs/1505.03018}{arXiv:1505.03018}.

\bibitem[Yoshida(2010)]{YoshidaOpenProblems2010}
R.~Yoshida.
\newblock Open problems on connectivity of fibers with positive margins in
  multi-dimensional contingency tables.
\newblock \emph{J. Algebr. Stat.}, 1\penalty0 (1):\penalty0 13--26, 2010.
\newblock \doi{10.18409/jas.v1i1.5}.

\bibitem[Yoshida and Barnhill(2023)]{Yoshida-Barnhill:Negative-cell}
R.~Yoshida and D.~Barnhill.
\newblock Connecting tables with allowing negative cell counts.
\newblock 2023.
\newblock Available at
  \href{https://arxiv.org/abs/2205.07167}{arXiv:2205.07167}.

\bibitem[Zhang and Chen(2013)]{ZhangChenJASA13}
J.~Zhang and Y.~Chen.
\newblock Sampling for conditional inference on network data.
\newblock \emph{Journal of the American Statistical Association}, 108\penalty0
  (403), 2013.

\end{thebibliography}
\bibliographystyle{abbrvnat}

\appendix

\section{Proofs}\label{appendix:proofs}

\subsection{Proof of Proposition \ref{thm:bounded graver is markov for bounded fibers}}
	Let $x$ and $y$ be two tables, in vector format, from the same restricted fiber $\mathcal F^q(b)$. The difference $y-x$ can be written as a conformal sum of primitive moves: \[y-x=z_1+\dots+z_r,\] where $z_i\in Gr(A)$ are elements of the Graver basis. The existence of such a decomposition follows from the definition of the Graver basis and is well-known in the literature (see for example \cite[section 4.6]{AHT2012}, ): the set of moves that cannot be written as a conformal sum of two nonzero moves is the Graver basis itself).  
	The definition of a `conformal sum' is that there is no cancellation of signs in the summands; in other words, cells in which values are added/subtracted by a move $z_i$ will not be subtracted/added by another move $z_j$. 
	
	Since there is no cancellation of signs in $z_1+\dots+z_r$, then $x+z_1+\dots+z_j$ belongs to the same fiber for each $j=1,\dots,r$, and none of the sums $x+z_1+\dots+z_j$ are outside the restricted fiber $\mathcal F^q(b)$. Because $y$ respects the cell bounds $\leq q$, and $x$ does too, it follows that in no sub-sum $x+z_1+\dots+z_j$  can one create an entry in a cell that is $>q$ because otherwise the sum is not conformal. \hfill $\square$

\subsection{Proof of Corollary \ref{cor:polynomial_hierarchical}} 

Let $\Delta$ be a simplicial complex with ground set $[k]$ and maximal faces $F_1, \ldots, F_r$. Let $V\subset [k]$ such that for every $i\in [k]$, $V\subset F_i$ or $V\subset F_i^c$. We will prove that , $A_\Delta=[A,B]^{(D_V)}$ where $A$ is a $\sum_{j:F_j\supseteq V}\frac{D_j}{D_V}\times \frac{D}{D_V}$ matrix,  $B$ is a $\sum_{j:F_j^c\supseteq V}D_j\times \frac{D}{D_V}$ matrix and $D_V=\prod_{l\in V}d_l$.

First notice that the columns of $A_\Delta$ are in bijection with the $D$ entries of a $d_1\times \cdots \times d_k$ table so we can label each column with an multi-index $i=(i_1, \ldots, i_k)$. For each multi-index $i$ and $F\subset [k]$ we define $i_F:=(i_j)_{j\in F}\in \prod_{l\in F}[d_l]$. Observe that each row can be identified with a pair $(F_m, f)$ where $f\in \prod_{j\in F_m}[d_j]$. Without loosing generality assume that $V=\{1, \ldots, v\}$ for some $v\in [k]$ and assume that $V\subset F_1, \ldots, F_s$ and $V\subset F_{s+1}^c, \ldots, F_r^c$. 

Now, following the construction of \cite{HoSu} we give a description of $A_\Delta$ as an $D_V$-fold matrix. Let us order the columns of $A_\Delta$ lexicographically: this order provides a partition of the columns into groups labeled by multi-indices in $\prod_{i=1}^v[d_i]$, i.e., the group corresponding to $i_V$ is the set $\mathcal C_{i_V}:=\{(i_V, i'): i'\in \prod_{l=v+1}^n[d_l]\}$, lexicographically ordered. 

The rows will be ordered in the following way: For each $i_V\in \prod_{l=1}^v[d_l]$ define the set of rows $\mathcal R_{i_V}:=\{(F_j, f): j\in [s], f_V=i_V\text{ and }f_l\in [d_l]\text{ for every }l\notin V\}$. Furthermore, given $(F_j, f), (F_{j'}, f')\in \mathcal R_{i_V}$ we say $(F_j, f)\prec (F_{j'}, f')$ if $j< j'$ or if $j=j'$ and $f$ is lexicographically smaller than $f'$. We denote by $\mathcal R$ the rest of the pairs $(F, f)$ that don't belong to any $\mathcal R_{i_V}$. Finally, we order the rows of $A_\Delta$ by groups $\mathcal R_{i_V}$ using a lexicographic order on $\{i_V:i\in \prod_{j=1}^k [d_j]\}$ and leaving the rows $\mathcal R$ at the end in any order. T

This order of the rows and columns provides a block description $[A, B]^{n}$ for $A_\Delta$ where $A=A_{link_\Delta(V)}$ and $B=A_{\Delta\minus V}$ are the design matrices of the hierarchical model associated to $link_\Delta(V):=\{F\minus V: F\supset V\}$ and  $\Delta\minus V=\{F\in \Delta: V\subseteq F^c\}$, respectively. \hfill$\square$

\subsection{Proof of Theorem \ref{thm:Unbounded_q}}

Let $n\geq 3$ and Consider the $(n-2)\times n$ integer matrix defined as follows:
\[
A_{n-2} = 
\begin{pmatrix}
1 & -2 & 1 & 0 &\cdots & 0 & 0 & 0\\
0 & 1 & -2 & 1 & \cdots & 0 & 0 & 0\\
\vdots & \vdots & \vdots & \vdots & \ddots & \vdots & \vdots &\vdots\\
0 & 0 & 0 & 0 &\cdots & -2 & 1 & 0\\
0 & 0 & 0 & 0 &\cdots & 1 & -2 & 1
\end{pmatrix}.
\]
In other words, we have that 
\[
A_{n-2}(i,j)=
\begin{cases}
    1, & \text{if } i=j \text{ or } j=i+2\\
    -2 & \text{if } j=i+1\\
    0 & \text{otherwise.}
\end{cases}.
\]
The column-style Hermite normal form of $A_{n-2}$ is given by $H=(I_{n-2}\; \bm 0_{n-2}\; \bm 0_{n-2})$ where $I_{n-2}$ is the $(n-2)\times(n-2)$ identity matrix and $\bm 0_{n-2}$ is the $(n-2)$-dimensional zero vector.

\vspace{.1in}

A simple computation shows that the $n\times n$ matrix
\[
U=
\begin{pmatrix}
    1 & 2 & 3 & \cdots & (n-1) & -(n-2)\\
    0 & 1 & 2 & \cdots & (n-2) & -(n-3)\\
    0 & 0 & 1 & \cdots & (n-3) & -(n-4)\\
    \vdots & \vdots & \vdots & \ddots & \vdots & \vdots\\
    0 & 0 & 0 & \cdots & 1 & 0\\
    0 & 0 & 0 & \cdots & 0 & 1
\end{pmatrix},
\]
is an unimodular matrix transforming $A$ into its column-style Hermite normal form, i.e, $AU=H$. Hence, it follows that the last 2 columns of $U$ provide a lattice basis for $\ker_\Z(A_{n-2})$. Let's denote by $L$ the $n\times 2$ matrix whose column vectors are the last two columns of $U$. Hence, as a consequence of (\cite{AHT2012}, Proposition 16.1)  we know that the column vectors of $\begin{pmatrix}L\\ -L\end{pmatrix}$ provide a lattice basis for $\Lambda(A_{n-2})$ where $\Lambda(A_{n-2})=\begin{pmatrix}A_{n-2} & 0_{n-2}\\ I_{n-2} & I_{n-2}\end{pmatrix}$ is the Lawrence lifting of $A_{n-2}$, being $0_{n-2}$ the $(n-2)\times (n-2)$ zero matrix. Let us denote the elements of this lattice basis by $z_1$ and $z_2$.

\vspace{.1in}

Now, consider the $n$-dimensional column vector $w=(0\; 1\; 2\; \cdots\; n-1)^T$ and let
\[
u=\begin{pmatrix} w\\  \bm 0_n\end{pmatrix}, \;\;\;\; v=\begin{pmatrix} \bm 0_n\\  w\end{pmatrix}.
\] 
Since $\Lambda(A_{n-2}) u=\Lambda(A_{n-2})v=(\bm 0_{n-2}\; w)^T$, we have that $u,  v$ are in the same fiber. However, adding any of the elements $\{\pm z_1,\pm z_2\}$ to $u$, results in at least one coordinate smaller or equal to $-(n-2)$. Hence, the lattice basis $\mathcal L_{n-2}:=\{z_1, z_2\}$ fails to connect $u,v$ inside $\mathcal F_{-q}((\bm 0_{n-2}, w)^T)$ for any $q=0, \ldots, n-3$. Therefore, for any $N>0$, $\Lambda_N:=\Lambda(A_{N+2})$ and $\mathcal M_N:=\mathcal L_{N+2}$ satisfy the properties stated in Theorem \ref{thm:Unbounded_q}.   
\hfill $\square$

\subsection{Proof of Theorem \ref{thm:bad-news_restricted}}

We begin by presenting the ideas used for the proof of Proposition \ref{prop:all-linear-are-transportation}. Let us start with a bounded polytope $P=\{y \in \R^n_{\geq 0}: Ay = b\}$,  where $A=(a_{i,j})$ is a $m\times n$ matrix. The construction of $T$ in \ref{prop:all-linear-are-transportation} is usually done in 3 steps (see \cite{DeLoera-Onn_all_linear}). However, for our purposes it will be enough to focus on the last 2 of them, listed below.
\begin{enumerate}[label=\textbf{\arabic*)}]
  \item[\textbf{Step 1)}] Representing $P$ as a plane-sum entry-forbidden transportation polytope $T'$.

  Let $U$ be an integer upper bound for the entries of $P$. Then, it can be proved that for some $s, h\in \Z^+$ and a subset $E\subset[s]\times [s]\times [h]$, $P$ can be represented as the polytope 
  \begin{align*}
  T'=\bigg\{ x \in \R^{s\times s\times h}_{\geq 0}: x_{i,j,k}=0 \text{ for all } (i,j,k)\notin E \text{ and } \\
  \sum_{i,j}x_{i,j,k}=c_k, \sum_{i,k}x_{i,j,k}=b_j, \sum_{j,k}x_{i,j,k}=a_i\bigg\}.
  \end{align*}
  This representation comes with an injection $\sigma':[n]\to [s]\times [s]\times[h]$ and its induced coordinate-erasing projection $\pi':\R^{s\times s\times h}\to \R^{n}$ that provides a bijection between $P$ and $T'$ and between their integer points.

  \begin{obs}\label{obs:embedding}
  From the description of $E$ in the proof of the result, it follows that for a given $y=\pi'(x)\in P$  (where $x\in T'$) and for any $i\in [n]$ the coordinate $y_i$ is embedded in $s_i$ distinct coordinates of $x$, where 
  \[
  s_i := \max \left(\sum_{k=1}^m\{a_{k,i}: a_{k,i}>0\}, \sum_{k=1}^m\{|a_{k,i}|: a_{k,i}<0\}\right).
  \]
  Moreover, the explicit set of coordinates where $y_i$ is embedded is given by 
  \begin{equation}\label{eqn:B_i-description}
  B_i := \bigg\{(j,j,\kappa(j)):j\in \Big\{1+\sum_{l<i}s_l, \ldots, \sum_{l\leq j}s_l\Big\}\bigg\}, 
  \end{equation}
  where $\kappa$ is a function $\kappa:\left[s\right]\to [h]$ completely determined by the matrix $A$ and $s=\sum_{i=1}^ns_i$ is the same as in the description of $T'$. Then, $B:=\bigcup_{i=1}^n B_i$ is a set of indices for which the corresponding entry in $x$ is equal to some $y_i$. Furthermore, the set $B$ is completely determined by $A$, so the previous embedding property holds for any $y \in P$.

  Notice that under the assumption that $A$ has nonzero columns we have that $|B|=\sum_{i=1}^ns_i$. In other words, the set $B$ is completely determined by the polyhedral representation of $P$. 
  \end{obs}

  \item[\textbf{Step 2)}] Representing the polytope $T'$ as a slim line-sum transportation polytope $T$.

  Given $T'$ as in the previous step, there are $r,c$ and $(u_{i,j})\in \Z^{I\times c}, (v_{i,k})\in \Z^{I\times 3}, (w_{j,k})\in \Z^{J\times 3}$ such that the transportation polytope 
  \[
    \widehat{T} = \left\{ x\in \R^{I\times J\times 3}_{\geq 0}: \sum_kx_{i,j,k}=u_{i,j}, \sum_jx_{i,j,k}=v_{i,k}, \sum_ix_{i,j,k}=u_{j,k}\right\}
  \] represents $T'$. 
\end{enumerate}  

The previous discussion leads to the following proof.

\begin{proof}[Proof of \ref{thm:bad-news_restricted}] Given a polytope $Q=\{ x \in \R^k_{\geq 0}: C x=  d\}$, and a vector ${ u=(u_1, \ldots, u_k)\in \Z^{k}}$ we let $Q_u:= \{ x \in \R^k: C x = d \text{ and } x_i\geq u_i \text{ for every }i\in [k]\}$. Also, given any $D\subset [k]$, we let $ \mathds{1}_D$ be the indicator vector of $D$ and write $\mathds{1}$ when $D=[k]$.

\vspace{.1in}

Now, consider the polytope $P=\{ y \in \R^{\eta+2}_{\geq 0}: y_0+y_{\eta+1}=1, \theta_jy_0-y_j=0, j =1, \ldots \eta\}$ introduced in the proof of Corollary \ref{cor:bad-news} and let $\widehat{P}:= P_{- 1}+ \mathds{1}$. The integer points in $\widehat P$ are exactly
\begin{align*}
 y^1 = (0,0,\ldots, 0,1)+ \mathds{1},\;\;\;  y^2 = (1, \theta_1, \ldots, \theta_\eta, 0)+ \mathds{1}, \;\;\;  \text{ and }\\   z^1 = (2, 2\theta_1, \ldots, 2\theta_\eta, -1)+ \mathds{1},\;\;\;  z^2 = (-1, -\theta_1, \ldots, -\theta_\eta, 2)+ \mathds{1}   
\end{align*}

By the previous discussion, there are $s, h\in \Z^+$ and a plane-sum entry-forbidden transportation polytope $\widehat T'\subset \R^{s\times s\times h}_{\geq 0}$ representing $\widehat P$. Furthermore, by \ref{obs:embedding} we know that there is a subset $B\subset [s]\times [s]\times[h]$ such that for any $ x\in \widehat T'$, the entries corresponding to the indices in $B$ are all entries of $\pi'( x)\in P$.

Let $r,c$ and $(u_{i,j})\in \Z^{I\times c}, (v_{i,k})\in \Z^{I\times 3}, (w_{j,k})\in \Z^{J\times 3}$ such that the transportation polytope 
\[
\widehat{T} = \left\{ x\in \R^{I\times J\times 3}_{\geq 0}: \sum_kx_{i,j,k}=u_{i,j}, \sum_jx_{i,j,k}=v_{i,k}, \sum_ix_{i,j,k}=u_{j,k}\right\}
\]
represents $\widehat T'$ and let $\sigma:[s]\times [s]\times [h]\to [r]\times[c]\times[3]$ be the injection given by this representation. Let $S=\sigma(B)$ and let $ p^1,  p^2,  q^1,  q^2\in \widehat T$ be the integer points corresponding to $ y^1,  y^2,  z^1,  z^2$, respectively. Then, consider the following transportation polytope 
\[
T = \left\{ x\in \R^{I\times J\times 3}_{\geq 0}: \sum_kx_{i,j,k}=u_{i,j}-(\mathds{1}_S)_{ij+}, \sum_jx_{i,j,k}=v_{i,k}-( \mathds{1}_S)_{i+k}, \sum_ix_{i,j,k}=u_{j,k}-(\mathds{1}_S)_{+jk}\right\}
\] 
and observe that $T_{- \mathds{1}_S} = \widehat T- \mathds{1}_S$. Moreover, since $\widehat T$ is a representation of $\widehat P$ it follows that the only integer points in $T_{- \mathds{1}_S}$ are $ p^1- \mathds{1}_S,  p^2- \mathds{1}_S,  q^1- \mathds{1}_S$ and $ q^2- \mathds{1}_S$. By construction, the first $2$ of these points are non-negative and any of the differences between any of the four points has either $\theta$ or $\theta$ appearing in the restriction of some $\eta$ coordinates. 

To see that $|B|=1+\sum_{i=1}^\eta\theta_i$ it is enough to find $|B_i|$ using \ref{eqn:B_i-description} and the defining matrix of the polytope $P$. 
\end{proof}

\subsection{Proof of Proposition \ref{prop:anti-stair}}

Before providing a proof let's introduce some notation. Given $i_1\neq i_2\in [I], k_1\neq k_2\in [3]$ and $j'\in [J]$ we define the $I\times J\times 3$ table $ b (i_1, i_2;\;j';\; k_1, k_2)$ as follows.
\[
b (i_1, i_2;\;j';\; k_1, k_2)_{i,j,k}=
\begin{cases}
  1, &\text{if } (i,j,k)\in \{(i_1, j', k_1), (i_2, j', k_2)\}\\
  -1, &\text{if } (i,j,k)\in \{(i_1, j', k_2), (i_2, j', k_1)\}\\
  0, &\text{otherwise.}
\end{cases}
\]  

We can think of this table as the embedding of a 2-way basic move in a 3-way table. Even though $b (i_1, i_2;\;j';\; k_1, k_2)$ has one non-zero 2-margin (so it is not a move), it will help us to describe some moves more easily. For instance, $b (i_1,i_2;\; j_1, j_2;\; k_1, k_2)$ can be written as a sum of two embedded $I\times 3$-moves: $$b (i_1,i_2;\; j_1, j_2;\; k_1, k_2)= b(i_1, i_2;\; j_1;\; k_1, k_2)+ b(i_2, i_1;\; j_2;\; k_1, k_2).$$

\begin{defn} Given a $I\times J\times 3$ table $m$ and a subset $S=S_I\times S_J\times S_K\subset [I]\times[J]\times[3]$. We define $m_S$ as the restriction of $m$ on $S$. Under this definition $m_S$ is a $|B_I|\times|B_J|\times|B_K|$ table.
\end{defn}

\begin{proof}[Proof of \ref{prop:anti-stair}] Suppose without losing generality that $\tau:[J]\to [3]$ is a surjective function such that $[I]\times[J]\times[3]\minus S=\bigcup_{j=1}^J\{(i,j,\tau(j)): i \in [I]\}$. 

Now, we will provide a way to find an infinite family of $(-q,S)$-extended fibers for which their non-negative $I\times J\times 3$ tables are not connected by basic moves.

Let $m = (m_{i,j,k})$ be a $I\times J \times 3$ table such that $m_{i,j,k}=0$ for every $(i,j,k)\notin S$ and for every $t\in[3]$ let $S_t=\{(i,j, k)\in S : \tau(j)=t\}=[r]\times \tau^{-1}(t)\times ([3]\minus \{t\})$. Notice that $S_1, S_2, S_3$ form a partition of $S$. Moreover, the support of any basic move that can be added to $m$ while preserving non-negativity constraint must be contained in some $S_t$. Otherwise, if $b$ is a basic move such that $\text{supp}(b)\cap S_t, \text{supp}( b)\cap S_{t'}\neq \emptyset$, it would follow that $m + b$ has a $-1$ for some entry in $[I]\times [J]\times [3]\minus S$ by a pigeonhole principle argument.

This implies that we can connect $m$ to another $I\times J \times 3$ table $m'$ (with basic moves) if and only if for every $t\in[3]$ we can connect $m_{S_t}$ and $m'_{S_t}$ with basic moves in their respective $(-q)$-extended fiber. In particular, if $m'$ is connected to $m$ we must have that $m_{S_t}$ and $m'_{S_t}$ are in the same fiber for every $t\in [3]$. In the rest of the proof, we will build $m$ and $m'$ such that the latter doesn't hold.

For every $t\in [3]$ pick some $j_t\in \tau^{-1}(t)$ and let us consider the move
\begin{align*}
n =  b(1,2;\; j_1;\;1,2)+ b(1,2;\; j_2;\;2,3)+ b(1,2;\; j_3;\;1,3)
\end{align*}

\noindent of degree $6$. By the choice of $j_1, j_2$ and $j_3$, we know that we can add $n$ to $m$ while preserving the non-negativity constraint. i.e., $m':= m+ n$ is a non-negative table in the fiber of $ m$. Moreover, by the definition of $n$ it follows that $m'_{S_1}=  b(1,2;\; j_1;\;1,2)+ m_{S_1}$ and therefore the 2-margins of $m_{S_1}$ and $m'_{S_1}$ are not the same (their $ik$-margins differ), contradicting our previous observations. Hence $m$ and $m'$ are not connected by basic moves and therefore the fiber of $m$ is not connected by basic moves.
\end{proof}

\end{document}